%% file: main.tex
\documentclass[a4paper,11pt]{article}
\pdfoutput=1
\usepackage{jcappub}
\usepackage{aas_macros}

\usepackage{bm}
\usepackage{color}
\usepackage{xcolor}
\usepackage{xspace}

\usepackage{tabularx}
\def\beq{\begin{eqnarray}}
\def\eeq{\end{eqnarray}}

\usepackage{soul,xcolor}
\setstcolor{red}

\usepackage{physics}
\usepackage{ifthen}

\usepackage{graphicx}% Include figure files
\usepackage{dcolumn}% Align table columns on decimal point
\usepackage{bm}% bold math
\usepackage{subfigure}
\usepackage[nointegrals]{wasysym}
\usepackage{verbatim}
\usepackage{color}
\usepackage{amsmath}
\usepackage[utf8]{inputenc}
\usepackage{orcidlink}

\usepackage{enumitem,amssymb}
\newlist{todolist}{itemize}{2}
\setlist[todolist]{label=$\square$}
\usepackage{pifont}

\input{priors.tex}
\newcommand{\Planck}{{\it{Planck}}}

\title{The Atacama Cosmology Telescope DR6 and DESI: Structure formation over cosmic time with a measurement of the cross-correlation of CMB Lensing and Luminous Red Galaxies}

\input{authors_kim.tex}

\emailAdd{jaejoonk@sas.upenn.edu}

\abstract{We present a high-significance cross-correlation of CMB lensing maps from the Atacama Cosmology Telescope (ACT) Data Release 6 (DR6) with luminous red galaxies (LRGs) from the Dark Energy Spectroscopic Instrument (DESI) Legacy Survey spectroscopically calibrated by DESI. We detect this cross-correlation at a significance of 38$\sigma$; combining our measurement with the \Planck~Public Release 4 (PR4) lensing map, we detect the cross-correlation at 50$\sigma$. Fitting this jointly with the galaxy auto-correlation power spectrum to break the galaxy bias degeneracy with $\sigma_8$, we perform a tomographic analysis in four LRG redshift bins spanning $0.4 \le z \le 1.0$ to constrain the amplitude of matter density fluctuations through the parameter combination $S_8^\times = \sigma_8 \left(\Omega_m / 0.3\right)^{0.4}$.  Prior to unblinding, we confirm with extragalactic simulations that foreground biases are negligible and carry out a comprehensive suite of null and consistency tests. Using a hybrid effective field theory (HEFT) model that allows scales as small as $k_{\rm max}=0.6~h/{\rm Mpc}$, we obtain a 3.3\% constraint on $S_8^\times = \sigma_8 \left(\Omega_m / 0.3\right)^{0.4} =  0.792^{+0.024}_{-0.028}$ from ACT data, as well as constraints on $S_8^\times(z)$ that probe structure formation over cosmic time. 
Our result is consistent with the early-universe extrapolation from primary CMB anisotropies measured by \Planck~PR4 within 1.2$\sigma$.  Jointly fitting ACT and \Planck~lensing cross-correlations we obtain a 2.7\% constraint of $S_8^\times = 0.776^{+0.019}_{-0.021}$, which is consistent with the Planck early-universe extrapolation within 2.1$\sigma$, with the lowest redshift bin showing the largest difference in mean.  The latter may motivate further CMB lensing tomography analyses at $z<0.6$ to assess the impact of potential systematics or the consistency of the $\Lambda$CDM model over cosmic time.}
 
\begin{document}
\maketitle
\setcounter{tocdepth}{1}

\tableofcontents

\section{Introduction}

The standard cosmological model, featuring cold dark matter (CDM) and a cosmological constant $\Lambda$, has been largely successful in describing how primordial density fluctuations developed into the present-day matter distribution. Our picture of the early universe is informed by the primary anisotropies in the cosmic microwave background (CMB) \cite{Planck_2018_VI, Rosenberg_2022, Aiola_2020, Balkenhol_2023}, which consists of radiation from the epoch of recombination at $z \approx 1100$. As these photons pass through gravitational potentials on their journey to us, they are deflected due to gravitational lensing (e.g., \cite{Lewis_2006}) allowing the CMB to be used as a probe of the late-time matter distribution as well. Together with complementary probes of the late universe including galaxy clustering \cite{Icaza-Lizaola_2019, 2112.04515, Simon_2023}, cluster cosmology \cite{ghirardini2024srgerosita, bocquet2024spt} and galaxy weak lensing \cite{Flaughter_2015, Abbott_2021, Kuijken_2015, Heymans_2020, Aihara_2017, More_2023, Miyatake_2023, Sugiyama_2023, Dalal_2023, Li_2023}, a suite of observables have reached the precision required to informatively compare with the prediction from early-universe CMB measurements. 

The matter distribution is typically characterized in terms of $\sigma_8$, the amplitude of matter density fluctuations smoothed on a scale of $8 \, \text{Mpc} / h$. Weak lensing observables, in particular, measure degenerate combinations with the average matter density of the universe $\Omega_m$, e.g., $S_8 = \sigma_8 \sqrt{\Omega_m / 0.3}$. Early observations of galaxy lensing with the CFHTLens survey \cite{10.1111/j.1365-2966.2012.21952.x} began to hint at a possible disagreement of this quantity between direct late-time observables and the primary CMB prediction \cite{1502.01589,1611.08606,10.1093/mnras/stw2665}.  Today, primary CMB measurements provide strong constraints on $S_8$ (derived through extrapolation to late times and assuming the $\Lambda$CDM model), e.g., $S_8 = 0.834 \pm 0.016$ from \Planck~2018 (PR3) \cite{Planck_2018_VI}, $S_8 = 0.827 \pm 0.013$ from \Planck~NPIPE (PR4) \cite{Rosenberg_2022}, $S_8 = 0.830 \pm 0.043$ from ACT DR4 \cite{Aiola_2020}, and $S_8 = 0.797 \pm 0.042$ from the South Pole Telescope (SPT-3G, \cite{Balkenhol_2023}) while measurements of the combination of galaxy weak lensing and galaxy clustering from surveys such as the Dark Energy Survey (DES, \cite{Flaughter_2015, Abbott_2021}), the Kilo-Degree Survey (KiDS, \cite{Kuijken_2015, Heymans_2020}), and the Hyper Suprime-Cam (HSC, \cite{Aihara_2017, More_2023, Miyatake_2023, Sugiyama_2023}) typically tend to find lower values, $S_8 = 0.776 \pm 0.017$, $S_8 = 0.765^{+0.017}_{-0.016}$, and $S_8 = 0.775^{+0.043}_{-0.038}$ respectively. Low inferences are also found in full-shape analyses of galaxy clustering from the Baryon Oscillation Spectroscopic Survey (BOSS, e.g., \cite{2112.04515,2302.04414}), but clustering from the full Sloan Digital Sky Survey (SDSS, \cite{Alam_2021}) that includes BOSS data as well as the joint reanalysis of galaxy weak lensing data from DES Y3 and KiDS-1000 \cite{Abbott_2023} find slightly higher values. Intriguingly, measurements of the CMB lensing power spectrum that best infer properties of structure at intermediate redshifts $0.5 < z < 5$ \cite{madhavacheril2023atacama} are in good agreement with the primary CMB: $S_8 = 0.831 \pm 0.029$ from \Planck~PR4 \footnote{The value of $S_8$ from \Planck~PR4 lensing was not explicitly provided in \cite{Carron_2022} but rather inferred from the chains provided in Section IV: \url{https://github.com/carronj/planck_PR4_lensing}} \cite{Carron_2022}, $S_8 = 0.840 \pm 0.028$ from ACT DR6 \cite{madhavacheril2023atacama,qu2023atacama} and $S_8 = 0.836 \pm 0.039$ from SPT-3G \cite{Pan_2023}. Galaxy cluster abundance measured by SPT (\cite{bocquet2024spt}) gives an intermediate value of $S_8 = 0.795 \pm 0.029$,  while an analysis using the first eROSITA All-Sky Survey (eRASS1, \cite{ghirardini2024srgerosita}) presents a higher value of $S_8 = 0.86 \pm 0.01$. 

Discrepancies between various probes could be sourced by systematics (e.g., unaccounted for baryonic feedback on small scales \cite{2206.11794, 2305.09827}), due to new physics (see e.g., \cite{2301.08361}), or caused by statistical fluctuations. Disentangling these requires observables across a range of redshifts and comoving wave-numbers, as well as observations that constrain feedback, e.g., \cite{2009.05557, 2009.05558}.  In this context, the cross-correlation of CMB lensing with the galaxy distribution can provide insight by exploring a wide range of redshifts while minimizing sensitivity to uncertainties on small scales. Recent galaxy-CMB lensing cross-correlation analyses show varying results: the cross-correlation of DES Y3 \verb|MagLim| galaxies with ACT DR4 CMB lensing \cite{Marques_2024} constrains $S_8 = 0.75^{+0.04}_{-0.05}$, the cross-correlation of BOSS with \Planck~PR3 \cite{Chen_2022} yields $S_8 = 0.707 \pm 0.037$, the cross-correlation of DES Y3 with SPT-SZ and \Planck~PR3 \cite{Chang_2023} presents $S_8 = 0.736^{+0.032}_{-0.028}$, while the cross-correlation of unWISE galaxies with the newer ACT DR6 CMB lensing and \Planck~PR4 \cite{farren2023atacama} shows $S_8 = 0.805 \pm 0.018$.

Cross-correlations with spectroscopically calibrated galaxy samples, in particular, have the potential to add significant additional robustness to tomographic studies.  The Dark Energy Spectroscopic Instrument (DESI) survey \cite{2022AJ....164..207D,2016arXiv161100037D,2023AJ....165....9S,2023arXiv230606310M,2016arXiv161100036D,2013arXiv1308.0847L,2023AJ....165..144G,2023AJ....166..259S} has collected $O(10^6)$ redshifts which we use here to calibrate the redshift distribution of target galaxies from the DESI Legacy Imaging Surveys \cite{Dey_2018}.  A previous \Planck~CMB lensing  cross-correlation analysis \cite{White_2022} used a similarly calibrated DESI sample and found a value of $S_8 = 0.73 \pm 0.03$ that is discrepant with the CMB prediction at $\sim 3 \sigma$.  In this work, we include lensing maps from the Atacama Cosmology Telescope (ACT) Data Release 6 (DR6), along with newer \Planck~CMB lensing maps from PR4 as well as several improvements to the analysis and theory modeling.

This paper is one of two papers along with \cite{Sailer_2024} analyzing the tomographic cross-correlation between ACT DR6 CMB lensing and the DESI luminous red galaxies (LRGs). In our companion paper \cite{Sailer_2024}, we delve into further details of the galaxy sample, discuss the HEFT model used in the analysis, and present constraints on $S_8$ and $\sigma_8$ when combining with baryon acoustic oscillation (BAO) data.  This paper details the methods and systematics in computing the galaxy-CMB lensing cross-correlation signal as an angular power spectrum and combines that with the DESI LRG auto-correlation angular power spectrum measurement to break the galaxy bias degeneracy. To demonstrate the constraining power of our analysis, this paper reports our best-constrained amplitude parameter $S_8^\times = \sigma_8 (\Omega_m/0.3)^{0.4}$ (with a slightly different exponent from $S_8$), including as a function of redshift. 

The outline of this paper is as follows: Section \ref{sec:data} discusses the CMB lensing and LRG data used in this analysis, Section \ref{sec:measurement} details the cross-correlation measurement computed with this data, Section \ref{sec:simulations} describes the generation and usage of simulations, including the calculation of the multiplicative transfer function in Section \ref{sec:transfer}, and the formulation of the analysis covariance matrix is described in Section \ref{sec:covmat}. Various null and consistency tests of our data and spectra are discussed in Section \ref{sec:nulls}. The cosmological parameter inference is described in Section \ref{sec:cosmology}, and finally, discussion of the results is presented in Section \ref{sec:results}.

\section{Data}

\label{sec:data}
\begin{figure}
\centering
\includegraphics[width=\columnwidth]{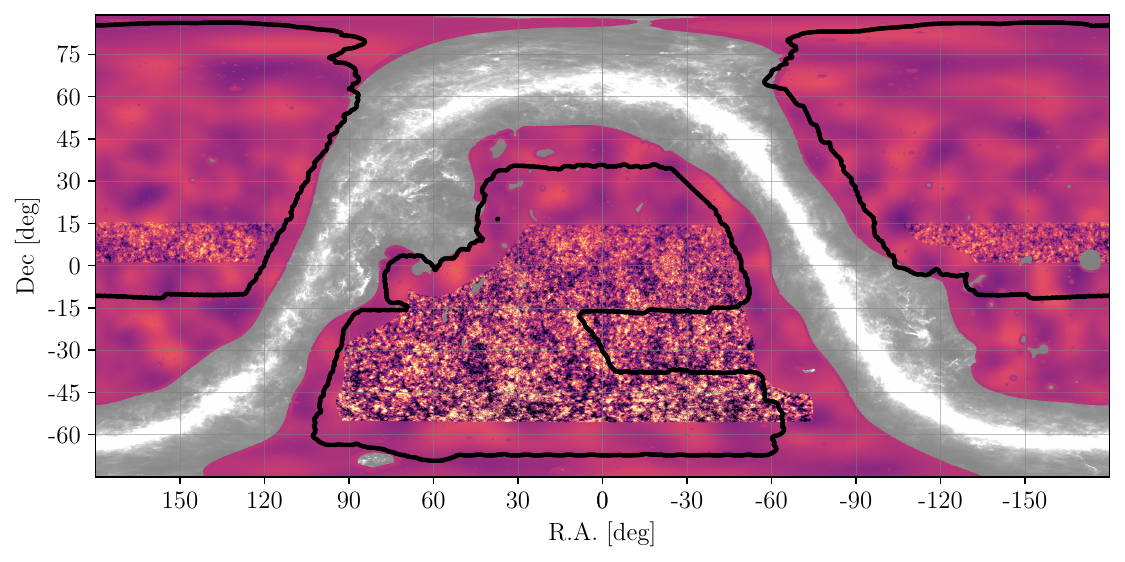}
  \caption{The Wiener filtered lensing convergence maps from \Planck~PR4 (blurry, background) and ACT DR6 (sharp, foreground) are shown here in equatorial coordinates, with the complete LRG footprint from DESI-LS shown as a black outline. The joint footprint between ACT and DESI spans approximately 19\% of the full sky (\Planck~and DESI cover $\approx$ 44\% jointly), with the mutually excluded region shown in gray surrounding the Galactic plane.}
    \label{fig:analysis_mask}
\end{figure}

In this work, we cross-correlate a sample of luminous red galaxies (LRGs) from the DESI survey with lensing mass maps from ACT DR6 as well as \Planck~PR4, with the respective footprints shown in Figure \ref{fig:analysis_mask}. In Section \ref{sec:desi}, we briefly summarize the properties of the galaxy sample from \cite{zhou2023desi,Zhou_2023} that is used in this analysis, and point the reader to the companion paper \cite{Sailer_2024} for further details. In Section \ref{sec:act} and Section \ref{sec:pr4}, we describe the CMB lensing data sets from ACT DR6 \cite{qu2023atacama}, \cite{madhavacheril2023atacama} and \Planck~PR4 \cite{Carron_2022} respectively and how they will be used for this analysis.

\subsection{DESI Luminous Red Galaxy sample}
\label{sec:desi}

The galaxy data used in this analysis is the ``Main LRG'' sample from \cite{zhou2023desi}, selected from DESI Legacy Imaging Surveys Data Release 9 (DESI-LS, DR9) photometric data with redshift distributions calibrated using the DESI Survey Validation (SV) dataset and Early Data Release \cite{DESI:2023dwi,DESI:2023ytc}. DESI-LS is an imaging survey to provide targets for DESI that consists of (a) galaxies lying north of declination $32.375^{\circ}$ sourced from observations by the Beijing-Arizona Sky Survey (BASS) of the Kitt Peak National Observatory and the Mayall $z$-band Legacy Survey of the Mayall Telescope, as well as (b) galaxies lying south of that declination covered by the Dark Energy Camera (DECam), with DECam providing imaging data to both the Dark Energy Camera Legacy Survey (DECaLS) and the Dark Energy Survey (DES). To see overlaps between imaging regions contributed from these different surveys, we refer the reader to Figure 2 of \cite{Sailer_2024} -- these regions combined lead to a total imaging area of 18,200 deg$^2$ after appropriate cleaning and masking steps.
 
\begin{figure}
\includegraphics[width=0.5\columnwidth]{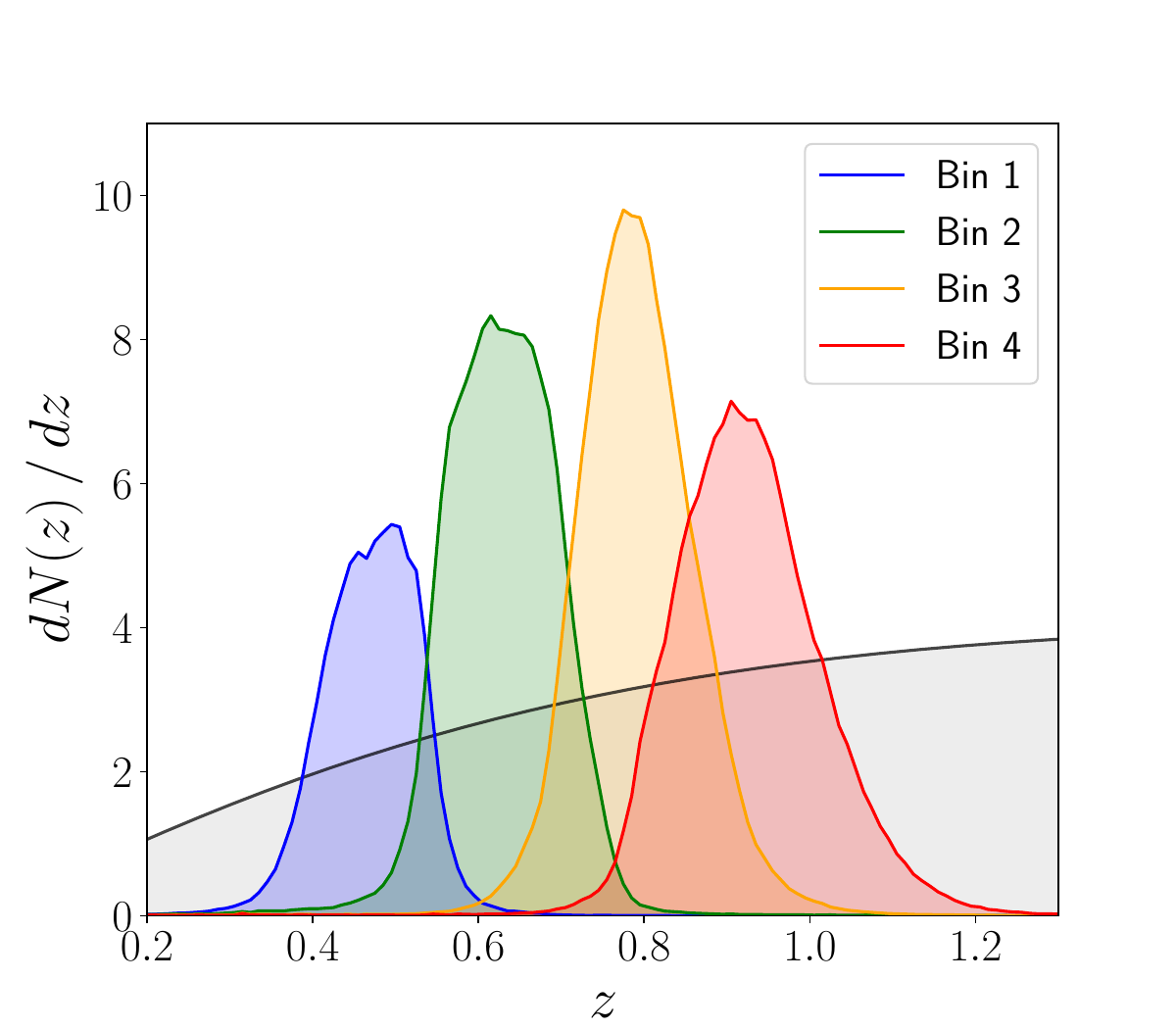}
\includegraphics[width=0.49\columnwidth]{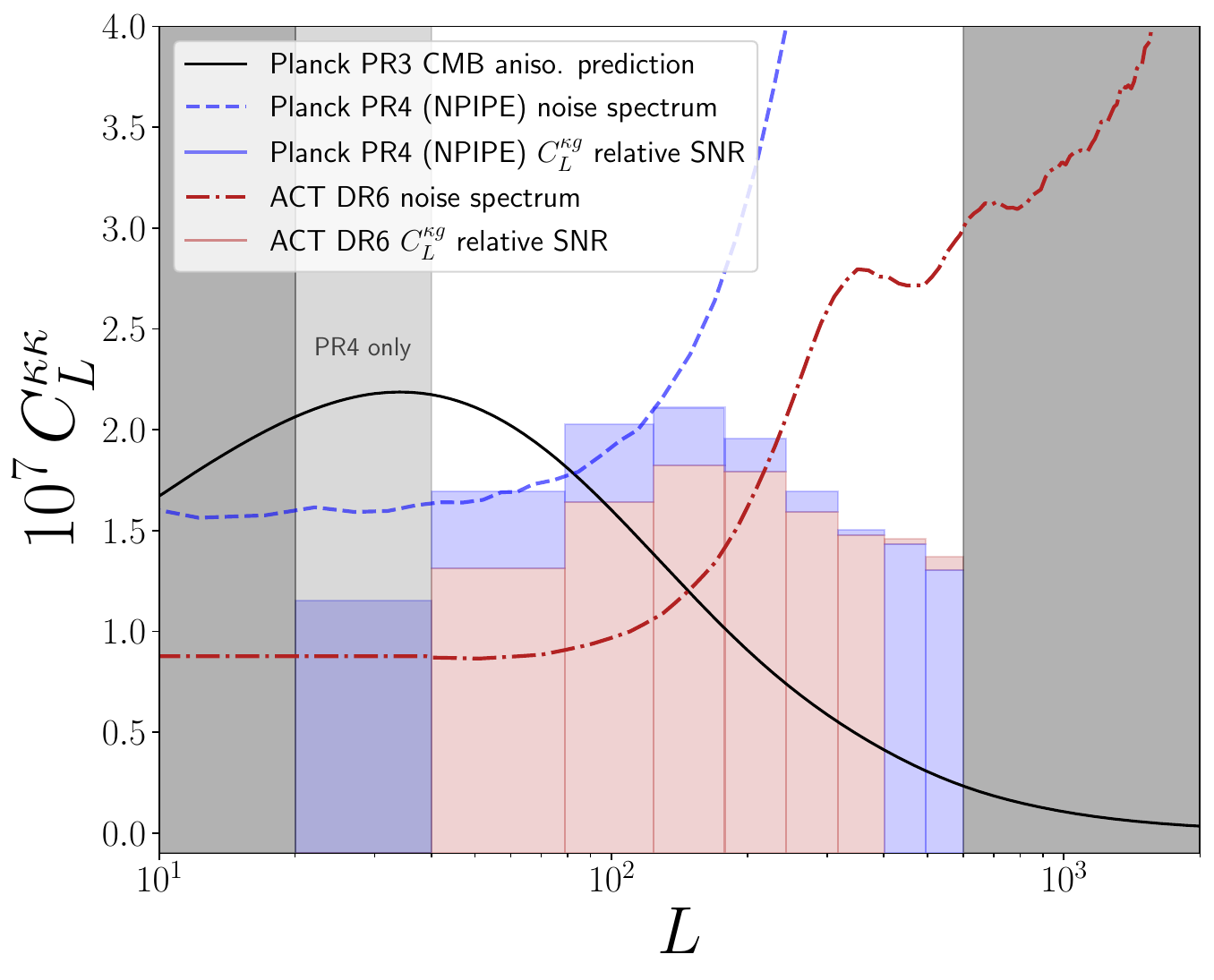}
  \caption{{\it Left: } The redshift distribution $dN(z)/dz$ of the DESI LRG galaxy redshift bins with the CMB lensing kernel shown in gray, showing ample overlap in redshifts between the two sets of cosmological probes. {\it Right: } \Planck~CMB prediction for the lensing power spectrum plotted against the lensing noise spectra of \Planck~PR4 (shown in blue) and ACT DR6 (shown in red). The lightly shaded bars in colors represent the fractional contribution to the cross-correlation $C_L^{\kappa g}$ signal-to-noise using covariances for \Planck~PR4 (blue) and ACT DR6 (red) and the same fiducial theory for both (see Appendix \ref{sec:snr-calculation} for more details), showing us that \Planck~holds more constraining power than ACT until $L \approx 400$. The shaded bars in gray show angular multipoles excluded due to scale cuts chosen for the analysis (where the light gray band labeled ``PR4 only" denotes an $L$ band included only for the \Planck~PR4 cross-correlation). {\it Both: } The colored bars and contours for both figures in addition to the gray CMB lensing kernel in the left figure are scaled to some arbitrary normalization factor for ease of visualization.}
  \label{fig:clkk}
\end{figure}

The ``Main LRG'' sample is selected and subdivided into four galaxy redshift bins by their photometric redshifts (photo-$z$) with criteria detailed in \cite{zhou2023desi} (e.g., total number density of around 550 $\text{deg}^{-2}$ for all redshift bins combined), but have redshift distributions calibrated with great precision by 2.3 million spectroscopic redshifts from DESI's SV and Year 1 data that are weighted and corrected for redshift failures (see \cite{Zhou_2023, Sailer_2024}). The photo-$z$ are computed using a random forest regression on training data from DESI spectroscopic redshifts, Sloan Digital Sky Survey's DR16, and a variety of other sources listed in Appendix B of \cite{zhou2023desi}. The redshift distributions of our bins are shown in Figure \ref{fig:clkk}.

Before overdensity maps are created, a series of quality cuts were applied to the galaxy catalog that lead to a cleaned sample with a redshift failure rate of approximately 1\% and a stellar contamination fraction of 0.3\% (further details can be seen in \cite{zhou2023desi, Sailer_2024}). As described in \cite{zhou2023desi}, multiplicative systematic weights for depth and seeing (in the $g, r, z$ bands) as well as an $E(B-V)$ correction for Galactic extinction \cite{Schlegel_1998} are estimated and applied to a catalog of random galaxies generated in the DESI footprint\footnote{Correlations between our $E(B-V)$ map and large-scale structure have been noted in \cite{zhou2023desi}; however, we investigate and observe in Figure 10 of \cite{Sailer_2024} that these correlations should have little to no impact in our analysis.}. Each of the galaxies in the four redshift bins as well as the randoms are then histogram-binned into a \texttt{HEALPix} map according to their coordinates, with the overdensity computed as the mean-subtracted galaxy counts map divided by the weighted random counts map. The resulting DESI galaxy overdensity map and binary mask for each redshift bin are provided without any modifications from Section 7 of \cite{zhou2023desi}\footnote{\url{https://data.desi.lbl.gov/public/papers/c3/lrg_xcorr_2023/v1/maps/main_lrg/}}.

\subsection{CMB Lensing}
\label{sec:cmb-lensing}
\subsubsection{ACT DR6}
\label{sec:act}
Our cross-correlation with DESI LRGs uses the baseline CMB lensing convergence map from ACT Data Release 6, a high-fidelity lensing mass map that covers approximately 23\% of the sky and overlaps with the DESI LRG analysis region over 19\% of the sky. This lensing mass map \cite{madhavacheril2023atacama} is generated from night-time CMB data collected over 2017 to 2021 with the Advanced ACTPol (AdvACT) receiver of the Atacama Cosmology Telescope in Cerro Toco, Chile \cite{Thornton_2016} at frequencies of approximately 97 GHz (denoted as \verb|f090|) and 149 GHz (denoted as \verb|f150|), as described in \cite{qu2023atacama}.  While ACT has collected data over roughly 44\% of the sky, the lensing analysis applies a further cut for Galactic contamination (restricting to the 60\% of the sky with the lowest dust contamination) that reduces the fiducial lensing sky coverage to 23\%. 

After isolating this 23\% sky region using an apodized mask, the \verb|f090| and \verb|f150| CMB intensity and polarization Stokes Q/U maps produced from multiple detector arrays are co-added with inverse-variance weights inferred from the noise properties of each array-frequency to produce spherical harmonic modes of the CMB temperature T as well as polarization E and B-modes \cite{qu2023atacama}. These are then Wiener-filtered and inverse-variance-filtered (in spherical harmonic space), retaining only CMB angular multipole modes in the range of $600 < \ell < 3000$ \cite{qu2023atacama}, with additional anisotropic cuts in 2D Fourier space that avoid contamination from ground pick up. The lower multipole cut of $\ell_{\rm min} = 600$ aims to mitigate contamination from Galactic dust \cite{Challinor_2018, Beck_2020} while the upper multipole cut of $\ell_{\rm max} = 3000$ mitigates extragalactic foreground contamination from the thermal and kinetic Sunyaev-Zel'dovich (tSZ/kSZ) effects, the Cosmic Infrared Background (CIB), and radio point sources.

The co-added and filtered maps are then passed through a quadratic estimator pipeline that reconstructs a map of the CMB lensing signal by exploiting the coupling of CMB multipole modes induced by lensing \cite{Hu_2002}. A simulation-based estimate of a `mean-field' additive bias is subtracted from this estimate to produce the final map \cite{qu2023atacama}. Since the pipeline uses a split-based cross-correlation estimator \cite{Madhavacheril2021Estimator} that uses multiple time-interleaved splits with independent instrument noise, the subtracted mean-field is immune to assumptions about the ACT instrumental noise. For cross-correlations in particular, this allows the scatter on large scales to be reliably predicted. In addition, while the lensing reconstruction normalization of the map is initially calculated analytically assuming isotropic filtering, a simulation-based multiplicative bias is also estimated to account for non-idealities like anisotropic filtering in Fourier space. These corrections can be as large as 10\% \cite{farren2023atacama} but are primarily dependent only on analysis choices, and thus can be robustly accounted for. The baseline map we use also implements profile hardening \cite{Namikawa_2013,Sailer2020ProfileHardening} to deproject mode-coupling signatures induced by objects that resemble tSZ clusters, which has been shown in \cite{Sailer2020ProfileHardening,Sailer:2022jwt,maccrann2023atacama} to mitigate the contamination from all known extragalactic foregrounds at current CMB noise levels. 

While the input CMB maps were filtered on scales of $600 < \ell < 3000$, the quadratic estimator reconstruction allows the estimation of lensing map modes at even lower multipoles due to how distortions in smaller scale CMB multipoles are caused by lensing at larger scales. The baseline ACT lensing map is provided over a multipole range of $2 < L < 3000$ \footnote{We follow the standard convention of using the symbol $L$ for lensing map multipoles and $\ell$ for input CMB map multipoles.}, but only modes greater than $L_{\rm min} = 40$ are deemed suitable based on the results of null and consistency tests regarding the influence of the mean-field \cite{qu2023atacama}.  The maximum reliable multipole in the map depends on the specific analysis (both from considerations related to foreground contamination as well as theory modeling); while this was $L_{\rm max} = 763$ for the CMB lensing auto-spectrum \cite{qu2023atacama}, we adopt a slightly lower maximum multipole of $L_{\rm max} = 600$. This choice is discussed briefly in Section \ref{sec:measurement} and in more detail in \cite{Sailer_2024}.

\subsubsection{\Planck~PR4}
\label{sec:pr4}
In order to obtain the best possible constraint on the amplitude of structure formation, we also cross-correlate the DESI LRG sample with the CMB lensing convergence map from the \Planck~satellite's Public Release 4 (PR4) \cite{Carron_2022}. This map covers a sky fraction of 65\% and overlaps with the DESI LRG analysis region over a sky fraction of 44\%. While the overlap region is twice as large as for the ACT map, the ACT maps have significantly lower noise, leading to a comparable signal-to-noise ratio for the cross-correlation with DESI LRGs (shown in Figure \ref{fig:clkk}). Our baseline constraint on structure formation includes cross-correlations with both the ACT and \Planck~lensing maps, with the \Planck~map contributing information primarily in the region not covered by ACT.

The \Planck~PR4 lensing map uses a quadratic estimator pipeline applied to CMB maps from the improved NPIPE re-processing of \Planck~High Frequency Instrument (HFI) data, where an additional $\approx 8$\% of CMB data (relative to \Planck~PR3) from satellite re-pointing maneuvers were included along with various improvements to data processing \cite{Planck_2020_LVII}. CMB multipoles of $100 \le \ell \le 2048$ are included in the reconstruction (with the maximum multipole motivated by the \Planck~beam) and result in a lensing map with modes reliable down to $L=8$. The quadratic estimator is run on an internal linear combination (ILC) of multi-frequency maps obtained using the SMICA algorithm \cite{Planck_2018_IV}. The use of ILC foreground cleaning along with the relatively low maximum CMB multipole makes this lensing map less susceptible to extragalactic foreground contamination, whereas in the ACT case, profile hardening was required for robustness against foregrounds.  Along with inhomogeneous noise filtering, the PR4 analysis also uses the Generalized Minimum Variance (GMV) quadratic estimator \cite{2101.12193},  a variant that performs a joint Wiener-filtering of the intensity and polarization maps that accounts for their correlation. Along with a post-processing step of Wiener-filtering the reconstructed lensing convergence maps, these choices make this analysis near-optimal and lead to an approximately ~10\% improvement of the signal-to-noise ratio (SNR) of the PR4 lensing power spectrum compared to the PR3 result, while per-mode improvements of the SNR can be as large as 20\%.

In the common sky area between \Planck~and ACT, the CMB lensing reconstructions from the two experiments are correlated. For lensing modes that are signal-dominated in both \Planck~and ACT (low-$L$), the correlation is large since it is primarily sourced by the sample variance of the underlying cosmic density modes. For noise-dominated modes at higher $L$, the correlation is smaller, but not zero. This is due to the fact that reconstruction noise is not just from CMB instrument noise (uncorrelated between experiments), but also from the random fluctuations of the primary CMB itself.  In order to perform a near-optimal analysis, we use the full available area from both the ACT and \Planck~maps, but fully account for their correlation in our simulation-informed covariance matrix, as described in Section \ref{sec:covmat}.

\begin{figure*}
\includegraphics[width=\textwidth]{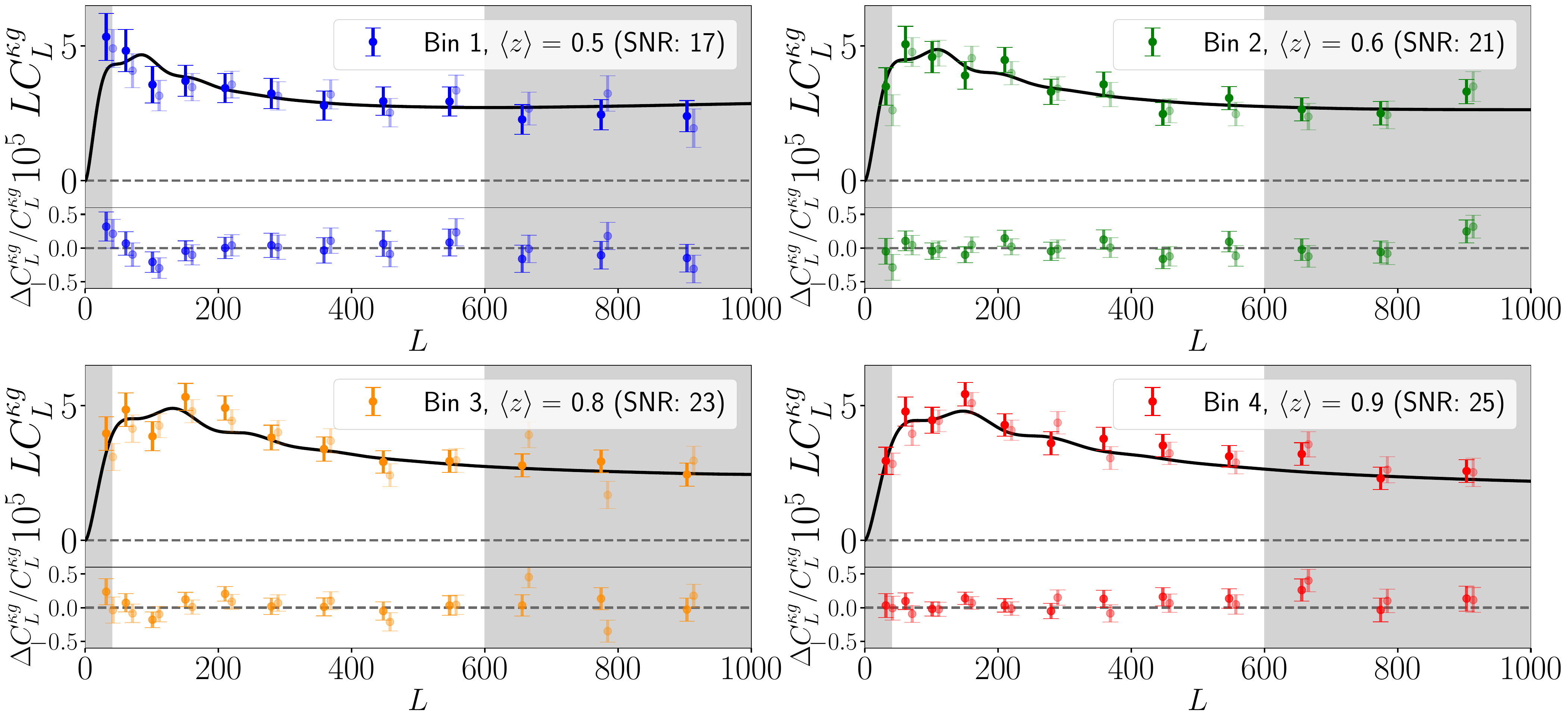}
  \caption{The ACT DR6 lensing x DESI LRG cross-correlation angular power spectra and residuals, for all four redshift bins, with the diagonal elements of their simulation-based covariances used for their respective error bars. The \Planck~PR4 x DESI LRG cross-correlation spectra are shown as lighter-shaded bandpowers that are slightly shifted to the right from the ACT bandpowers for visual purposes.  The signal-to-noise (SNR) ratio for each redshift bin is computed over the analysis $L$ range up to $L_{\rm max} = 600$. The solid black curve in each plot is the power spectrum computed from the fiducial model using baseline best-fit cosmological parameters jointly fit to all four redshift bins, their auto-spectra, and their cross-correlations with ACT and \Planck, within their respective analysis $L$ ranges. The best-fit spectra fit to 66 total degrees of freedom (computed from subtracting the number of free parameters of the model fit from the total number of bandpowers being fit to, henceforth ``d.o.f'') results in a $\chi^2 = 54.1$ (15 d.o.f for $\chi^2 = $ 11.5, 9.86, 16.1, 12.8 for each redshift bin fit independently). Assuming each free parameter removes exactly one degree of freedom, this leads to a probability-to-exceed (PTE) of 85.2\%, demonstrating a good fit; \cite{Sailer_2024} discusses the violation of this assumption for the case of prior-dominated parameters and provides a model fit PTE calculation.}
    \label{fig:clkg}
\end{figure*}

\section{CMB lensing tomography measurement}
\label{sec:measurement}
In spherical harmonic space, we perform an analysis of the two-point cross-correlation between the CMB lensing and the LRG overdensity fields as well as the two-point auto-correlation of the LRGs themselves. To constrain cosmology and the evolution of structure, we use a technique to use varying redshift slices of galaxies in computing these two correlations jointly known as \textit{CMB lensing tomography} \cite{Hu_1999}. In this section, we describe the formalism for measuring the angular power spectra and its implementation. We use this implementation to measure power spectra for our data products as well as simulations which we use to estimate a transfer function and the data covariance.

The cross-correlation between the CMB lensing convergence and the galaxy overdensity field can be expressed (under the Limber approximation \cite{Limber1953, 2008PhRvD..78l3506L})  as an integral over the line-of-sight comoving distance $\chi$ of the three-dimensional matter power spectrum, weighted by the CMB lensing and galaxy projection kernel functions $W^\kappa$ and $W^g$:
\begin{equation}
C^{\kappa g}_L = \int \, \dfrac{d\chi}{\chi^2} \, W^{\kappa}(\chi) W^{g}(\chi) P_{mg} \left(k = \dfrac{L + 0.5}{\chi}, z(\chi) \right).
\end{equation}

While the galaxy-matter cross-spectrum $P_{mg}(k)$ is proportional to the square of the amplitude of structure formation, it is also dependent on how galaxies trace the underlying matter density. To break this galaxy bias degeneracy, we also measure the auto-spectrum of the galaxy overdensity, which under the Limber approximation is:
\begin{equation}
    C^{gg}_L = \int \, \dfrac{d\chi}{\chi^2} \, W^{g}(\chi) W^{g}(\chi) P_{gg} \left(k = \dfrac{L + 0.5}{\chi}, z(\chi) \right)
\end{equation}
which is evaluated using the galaxy kernel function previously mentioned.

%\footnotetext{For each of the 4 redshift bins, there are 9, 8, and 7 bandpowers present for the \Planck-only cross-spectrum, ACT-only cross-spectrum, and the DESI LRG auto-spectrum respectively, resulting in 96 total data points minus 30 free parameters for the theory modeling (7 nuisance parameters per bin and 2 cosmological parameters, see Table \ref{tab:priors}) for a grand total of 66 degrees of freedom for our theory model's best-fit spectra fit to all 4 redshift bins simultaneously. For the case where we fit the theory model to a singular redshift bin, we have 24 data points minus 9 free parameters for modeling, resulting in 15 total degrees of freedom.}

Here $W^g$ encodes the redshift distribution of the LRGs and $W^\kappa$ the redshift dependence of contributions to the CMB lensing map \cite{1311.6200} (see Figure \ref{fig:clkk}). In practice, the above equations include additional terms to account for magnification bias \cite{bartelmann2001291} arising from the modulation of galaxy number counts by foreground lensing, and the 3D power spectra are built from an effective field theory (EFT) formalism: see Section \ref{sec:theory} here and Section 4.5 of our companion paper \cite{Sailer_2024} for additional details.   

\begin{figure*}
\includegraphics[width=\textwidth]{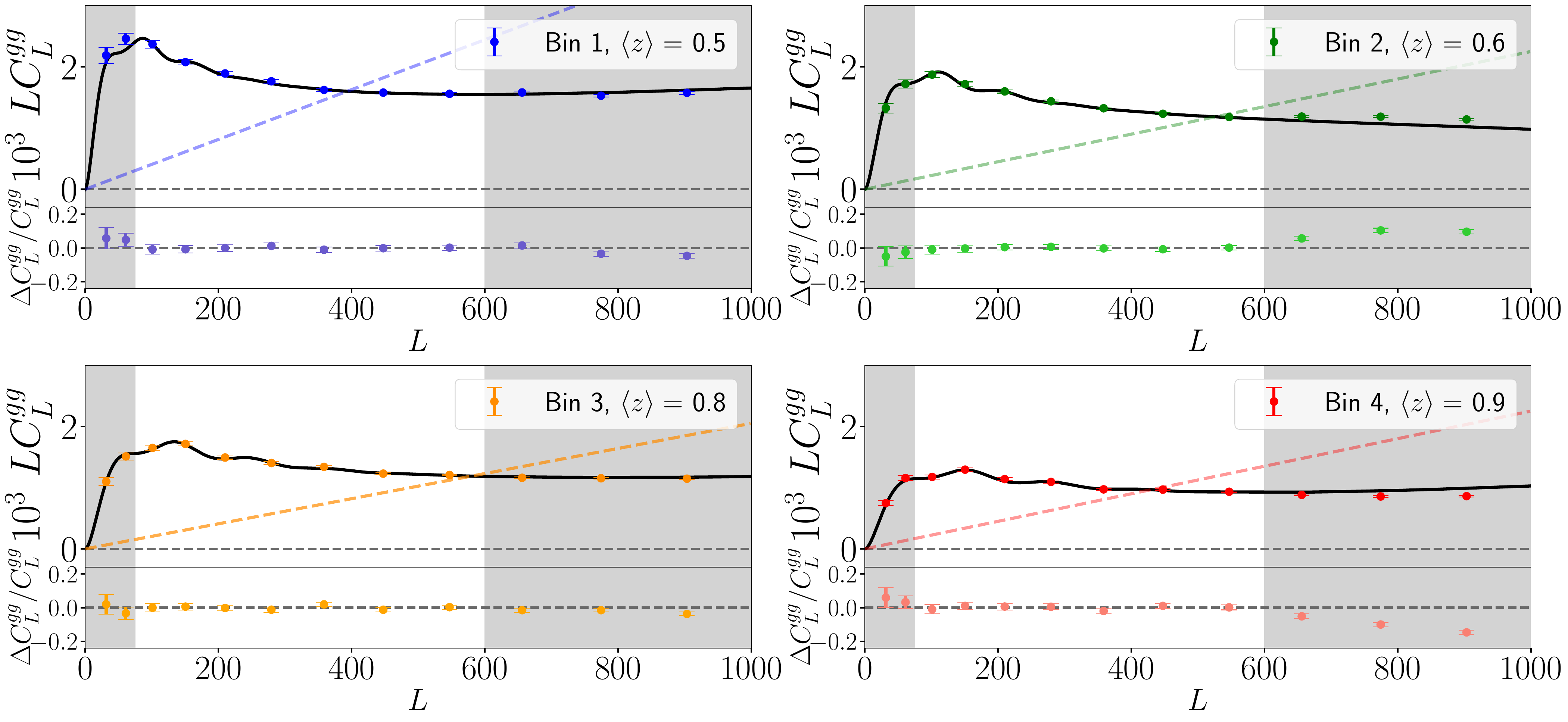}
  \caption{The DESI LRG angular auto power spectrum, with all four redshift bins and the diagonals of their simulation-based covariances used for their respective error bars. A fiducial value of the shot noise level estimated using a HEFT best-fit is subtracted for all four redshift bins, and is shown as colored dashed lines for the respective redshift bin. The power spectrum computed from the model described in the caption of Figure \ref{fig:clkg} (once again, fitting only to data in the non-gray regions) is shown in black; as demonstrated by the $\chi^2$ computation in Figure \ref{fig:clkg} ($\chi^2 = 54.1$, PTE = 85.2\%) this is indeed a good fit.}
    \label{fig:clgg}
\end{figure*}

The degeneracy between the galaxy bias model and the amplitude of structure formation is broken due to $C_L^{\kappa g}$ and $C_L^{gg}$ having different dependencies on the galaxy bias while both being proportional to $\sigma_8^2$, therefore a joint fit to the galaxy auto-spectrum and the galaxy-CMB lensing cross-spectrum allows us to constrain the growth of structure independently of the galaxy bias. We show our measurement for $C_L^{\kappa g}$ in Figure \ref{fig:clkg} and $C_L^{gg}$ in Figure \ref{fig:clgg}. In Section \ref{sec:cosmology} and Section 4 of \cite{Sailer_2024}, we discuss how our theory model accounts for non-linearities in galaxy biasing as well as the underlying matter power spectrum. 

\subsection{Angular power spectrum}
\label{sec:angular-power-spectrum}
A naive estimator for the angular power spectrum of two fields $X$ and $Y$ is:
\begin{equation}
    \tilde{C}_L^{XY} = \dfrac{1}{2L+1} \displaystyle\sum_{M = -L}^L x_{LM} y^*_{LM}
    \label{eqn:power-spectrum}
\end{equation}
in terms of the spherical harmonic decomposition of $X$ and $Y$ into coefficients $x_{LM}$ and $y_{LM}$, but care must be taken to account for mode-coupling introduced by masking and the inhomogeneous weighting of the maps. To compute an unbiased estimate of the angular power spectrum of two masked fields, we use the MASTER algorithm as detailed in \cite{Hivon_2002} and implemented by the \verb|NaMaster| code \cite{Alonso_2019}. The MASTER algorithm inverts the following relation between the biased power spectrum of the masked fields (pseudo-$C_L$, denoted as $\tilde{C}_L$) and the unbiased angular power spectrum $C_L$ using a mode-coupling matrix $M_{L L'}$ computed from the spherical harmonic coefficients of the masks:
\begin{equation}
    \label{eqn:mode-coupling}
    C_L^{XY} = \displaystyle\sum_{L'} M_{L L'} \tilde{C}_{L'}^{XY}.
\end{equation}
Due to the information loss caused by masking, the $L$-by-$L$  inversion of the mode-coupling matrix for a masked field is not possible; thus it is common to bin the coupled pseudo-$C_L$ into bandpowers with a set of normalized weights $\displaystyle\sum_{L=0}^{L_{\rm max}} w_L^b = 1$ for each bandpower bin denoted by ${L_b}$. Under the assumption that the underlying power spectrum is piecewise constant in each bin, these bandpowers can then be approximately decoupled using the inverse of the binned mode-coupling matrix, formulated by applying the same normalized weights $w_L^{b}$ to the mode-coupling matrix \cite{Alonso_2019}. The combination of bandpower weights and coupling matrix is accessed by \verb|NaMaster|'s bandpower window functions and specified by the binning scheme and mask geometries. 

To prepare an $L$-dependent function (such as a theory spectrum) $C'_L$ to compare directly with our estimation of the unbiased, binned angular power spectrum $C_{L_b}$, we convolve $C'_L$ with our bandpower window functions, which applies the coupling, binning, and decoupling steps altogether; this procedure can be different from naively binning $C'_L$ as the bandpower window functions correct for piecewise constant bins. The same procedure is used to evaluate the likelihood for our analysis to compare our binned angular power spectrum data vector with a $C'_L$ prediction from our theory model. For all purposes in this paper, the true angular power spectrum is computed by using the \verb|compute_full_master| method in \verb|NaMaster| that implements this pseudo-power spectrum estimator. 

The ACT DR6 lensing analysis mask is provided in \verb|HEALPix| pixelization format with \verb|Nside = 2048|, in the same format as the DESI LRG map and analysis mask. The ACT DR6 and \Planck~PR4 lensing convergence maps are provided as spherical harmonic coefficients that are first low-pass filtered to exclude $L > 3000$ and then transformed into \verb|HEALPix| maps of the same format. As all \Planck~data products are provided in Galactic coordinates while the ACT DR6 and DESI data products are in equatorial coordinates, we decompose the \Planck~PR4 mask into spherical harmonic coefficients, rotate the mask and map coefficients from Galactic to equatorial coordinates, and then transform them back into maps; this specific order keeps the power spectrum invariant between coordinate systems. Since the ACT DR6 lensing analysis mask is an apodized (non-binary) map that has effectively been applied twice during the process of lensing reconstruction through a quadratic estimator, we pass the square of the ACT lensing mask into the \verb|NaMaster| mode-coupling calculation as an approximation to account for this effect.

The mode-coupling inversion for a mask that has been applied before the use of a quadratic estimator is not exact, so we correct our \verb|NaMaster| power spectrum result by applying a simulation-based multiplicative transfer function (described in Section \ref{sec:transfer}). After computing the galaxy-CMB lensing cross-spectrum measurement, we used the exact same pipeline to iterate and cross-correlate the appropriate lensing simulations and their respective correlated Gaussian galaxy fields to aid in computing the covariance matrix elements (see Section \ref{sec:covmat} for more details).

Here, we have omitted the treatment of the scale-dependent pixel window function, which captures the effect of pixelizing a continuous two-dimensional sky map and remains to be accounted for when binning a catalog into a discretely pixelized map. This pixel window function, contributing approximately an order of a percent in the analysis scale range of this work, is in fact not corrected at the spectrum level and is instead forward-modeled for the likelihood (see Section \ref{sec:theory} and \cite{Sailer_2024} for further details); this is because the pixel window function correction for a galaxy sample's auto-spectrum requires it to be shot-noise subtracted. Instead, we proceed with a more assumption-agnostic, forward model approach of analytically marginalizing over the shot noise level, which allows us to model a pixel window-convolved result with our likelihood's theory predictions to compare directly with our data's cross-correlation bandpowers. A promising avenue for future iterations to this analysis is the method presented in \cite{lizancos2024harmonic} that computes angular power spectra by bypassing the usage of map pixelization and therefore, treatment of various systematics including harmonic-space aliasing, shot noise, and pixel window functions.

For the \Planck~PR4 cross-correlation measurement needed for the joint covariance, the analysis mask used for the lensing measurement is apodized with a $0.5^{\circ}$ \verb|C2| \footnote{As described in \cite{White_2022}, in terms of the angle from a masked pixel $\theta$ and the apodization angular scale $\theta^*$, the \texttt{C2} filter is a factor $f = 0.5 \left(1 - \cos \pi x \right)$ for $x = \sqrt{(1 - \cos\theta) / (1 - \cos\theta^*)}$ applied to all pixels for which $x < 1$. This data-based choice of apodization angular scale used in our analysis was adopted from \cite{White_2022}.} filter and is reapplied onto the PR4 lensing convergence map while performing a similar pseudo-$C_L$ computation routine with the same LRG footprint mask and maps. Since our pipeline manually apodizes the PR4 analysis mask and alters it from the binary mask used in the GMV lensing reconstruction, the power spectrum is computed with a re-application of one power of the PR4 lensing analysis mask (as opposed to the two powers used for the ACT DR6 lensing analysis mask) onto the lensing convergence map. The harmonic multipole range and format of the coupling matrix is the exact same as the one used for the ACT DR6 cross-correlation measurement. However, the transfer function applied onto this measurement is computed instead with 480 \Planck~PR4 lensing simulations that have been lensed from the FFP10 input lensing potentials (as described in \cite{Planck_2018_III}) with the appropriate footprint mask accounted for.

For all measurements, the bandpowers are binned by angular multipole intervals that are linear in $\sqrt{L}$, so our bins are computed as follows:
\begin{align*}
\text{Bin edges} &= [10, 20, 44, 79, 124, 178, 243, 317, \\
&\hphantom{= [}  401, 495, 600, 713, 837, 971, ...].
\end{align*}
All bandpowers, covariance matrices, and window functions are computed from an $L_{\rm min} = 10$ up to $L_{\rm max} = 6000$, but only used from $L'_{\rm min} = 20$ to $L'_{\rm max} = 1000$ to evaluate the likelihood in order to prevent any mode-coupling related power leakage near the multipole limits. Based on the $L_{\rm min}$ values discussed in Section \ref{sec:cmb-lensing}, we devise an analysis $L$-range for the galaxy-CMB lensing cross spectrum with ACT DR6 to range from $L_{\rm min} = 44$ to $L_{\rm max} = 600$ while with \Planck~PR4, we include the lowest analysis $L$ bin down to $L_{\rm min} = 20$. We choose $L_{\rm max} = 600$ that corresponds to the comoving distance to the peak of bin 1's redshift distribution with a $k_{\rm max} = 0.5 \, h/{\rm Mpc}$\footnote{This is equivalent to the $L_{\rm max}$ computed by using the lower edge of our lowest redshift bin with a $k_{\rm max} = 0.6 \, h/{\rm Mpc}$, the method described in the companion paper \cite{Sailer_2024}.} that is validated according to our theory model; this is ultimately a conservative choice as we apply the same scale cut to all other (higher) redshift bins. The galaxy auto-spectrum for the DESI LRGs is computed from $L_{\rm min} = 79$ instead, to circumvent the need to apply percent-level corrections to the Limber approximation due to redshift-space distortions \cite{Fisher_1993, Padmanabhan_2007}. This binning scheme allows consistency in computing all three sets of measurement bandpowers while being able to fully explore the angular scales available with our theory modeling and noise constraints. It also takes advantage of the idea that our signal-to-noise improvements are nominal at the smallest scales while being able to efficiently compress our data vectors and covariances, so we utilize sparser small-scale bandpowers while comprehensively capturing the signal amplitude at the largest scales.

\subsection{Simulations}
\label{sec:simulations}
To characterize multiplicative transfer functions and inform covariance matrices for correlations within and across data-sets, we build simulation suites that contain $\mathcal{O}(100)$ Gaussian realizations of the CMB, lensing reconstructions of the CMB, and correlated Gaussian random fields that are generated with a constraint of matching the power of a given fiducial power spectrum to represent a biased tracer of large-scale structure.

We start with Gaussian realizations of the CMB lensing convergence field $\kappa$ available from \cite{qu2023atacama,madhavacheril2023atacama}. From these, we generate correlated, simulated DESI LRG overdensity maps assuming some fiducial cross- and auto-spectra with CMB lensing. Specifically, as done in e.g., \cite{farren2023atacama}, we split the galaxy overdensity into a part correlated with CMB lensing and a part that is uncorrelated:
\begin{align}
    g_{LM} &= g_{LM}^{\rm corr.} + g_{LM}^{\rm uncorr.} \\
    \label{eqn:sim-gen}
    g_{LM}^{\rm corr.} &= \kappa_{LM} \times \dfrac{C_{L}^{\kappa g}}{C_{L}^{\kappa \kappa}} \\
    \left\langle g_{LM}^{\rm uncorr.} \left(g_{LM}^{\rm uncorr.}\right)^* \right\rangle  &= C_L^{gg} - \dfrac{(C_L^{\kappa g})^2}{C_L^{\kappa \kappa}}.
    \label{eqn:sim-gen-uncorr}
\end{align}

Each overdensity map is then a sum of the two components, with the correlated part being a re-scaled version of the CMB lensing convergence map and the uncorrelated part a new random realization drawn from the spectrum given by Eq. \ref{eqn:sim-gen-uncorr}; the correlated and uncorrelated parts represent the mean and variance terms respectively of a conditional distribution of drawing $g_{LM}$ given $\kappa$, where $g_{LM}$ and $\kappa$ are correlated Gaussian random variables of zero mean and some variance. It follows then that the power spectra computed using $g_{LM}$ agree with the fiducial prediction for both the galaxy auto-spectrum $C_L^{gg}$ and the cross-spectrum $C_L^{\kappa g}$ when ensemble averaged over all realizations.
When estimating transfer functions or covariance matrices using these simulations, we draw up to 10 Gaussian galaxy simulations for each lensing convergence simulation to reduce the noise on these estimates, noting that the choice of ten draws (in lieu of one draw) would decrease the correlation of our lensing simulations to noise and therefore our scatter on the simulated $C_L^{\kappa g}$ measurement. To compare directly to a data measurement of the galaxy power auto-spectrum $\tilde{C}_L^{gg}$ that includes the Poisson shot noise level $\tilde{N}_L^{gg}$, we compute $g_{LM}$ using the shot noise subtracted fiducial galaxy power auto-spectrum $\tilde{C}_L^{gg}$, and add back a \texttt{HEALPix}-formatted random white noise realization commensurate with the expected shot noise level. 

The ACT DR6 lensing suite comes with a set of 400 CMB simulations that are lensed by the Gaussian lensing convergence realizations used above that match a fiducial lensing auto power spectrum $C_L^{\kappa \kappa}$. The suite also provides 400 simulations for each of the reconstructed ACT DR6 lensing products, including ACT DR6 lensing reconstructions done on a null combination of CMB maps (e.g., a difference of the CMB mapped at different frequencies) and ACT DR6 lensing reconstructions done on variants of the maps (e.g., polarization only, curl component of the lensing field). As described in \cite{atkins2023atacama}, noise simulations with the ACT DR6 CMB noise levels are used alongside these simulations and passed through the lensing reconstruction pipeline described in \cite{qu2023atacama} to generate a reconstructed lensing simulation for each input CMB field. The iteration of cross-correlations over these 400 reconstructed lensing simulations with their correlated galaxy fields allows us to estimate a galaxy-CMB lensing cross-spectrum covariance for various null tests, the uncertainty in the transfer function, as well as the measurement bandpowers themselves.

Similarly, the \Planck~PR4 lensing suite comes with a set of 480 CMB simulations from FFP10 \cite{Planck_2018_III} that are lensed by independent Gaussian lensing potential realizations matching the lensing power auto-spectrum of a provided fiducial theory spectrum. As discussed previously in Section \ref{sec:measurement}, the \Planck~PR4 lensing simulations are rotated to equatorial coordinates, and their corresponding correlated galaxy fields are drawn from these simulations using Equation \ref{eqn:sim-gen-uncorr} to estimate the covariance for the \Planck~PR4 cross-correlation. These 480 FFP10 CMB simulations can also be used to generate lensing reconstructions correlated with both ACT and \Planck; in \cite{qu2023atacama}  and \cite{farren2023atacama}, an independent set of simulations was created by taking these lensed CMB realizations, masking them with the ACT DR6 analysis mask, and reconstructing their lensing convergence using the ACT DR6 lensing pipeline (using the same CMB angular scale cuts and other various lensing power spectrum analysis choices, while excluding instrumental noise). As mentioned before, since these output reconstruction simulations estimate similar lensing signatures from the same CMB fields using different analysis choices and pipelines, they are used to estimate correlations between the ACT DR6 and \Planck~PR4 lensing fields and their individual cross-correlations with DESI LRGs for a joint covariance matrix and correlated analysis.

\subsection{Transfer function}
\label{sec:transfer}

Following an in-depth discussion in \cite{farren2023atacama}, we estimate transfer function corrections to our cross-spectra for two main reasons: (a) the mode coupling deconvolution in the MASTER algorithm assumes that the mask has been applied at the level of the input field; however CMB lensing maps are produced from quadratic combinations of masked CMB fields and (b) to account for small spatially dependent normalization offsets in the lensing maps.

The latter are due to analysis choices in lensing reconstruction resulting in small levels of misnormalization in the map. For example, the ACT pipeline uses 2D Fourier space filtering whereas the analytic normalization of the estimator assumes isotropy. This leads to a 10\% mis-normalization, which is corrected in \cite{qu2023atacama} at the lensing map level through a simulation-based transfer function. That correction, however, is estimated on the full footprint of the ACT lensing map. The relevant correction for our cross-correlation analysis may be slightly different since the overlap with DESI selects a slightly smaller region of the ACT lensing map. Similarly, the \Planck~PR4 lensing analysis applies inhomogeneous filtering and corrects for the departure from analytic normalization using a simulation-based transfer function. Here too, we estimate an additional transfer function relevant to our cross-correlation in the DESI overlap region.

\begin{figure}
\includegraphics[width=0.49\columnwidth]{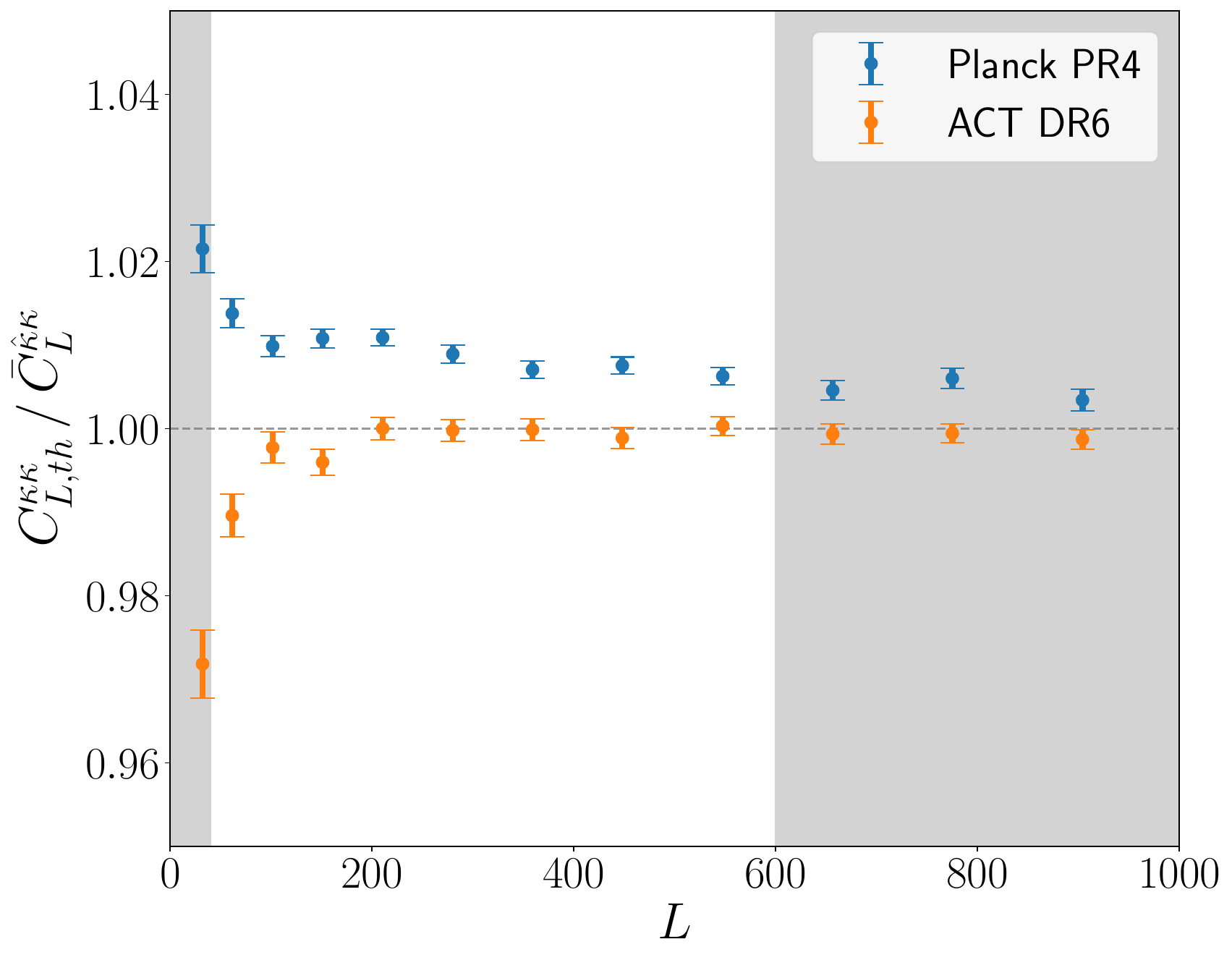}
\includegraphics[width=0.51\columnwidth]{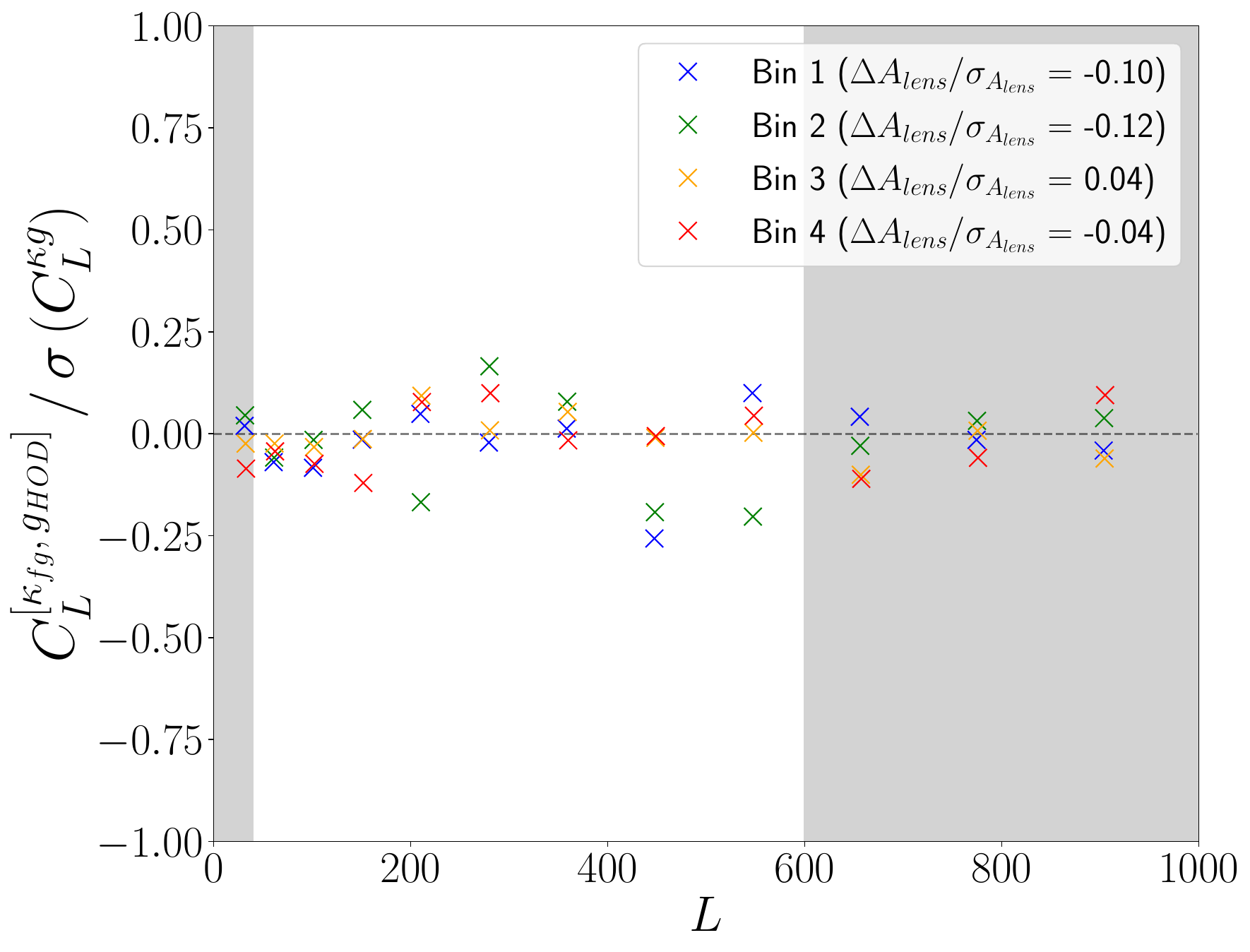}
  \caption{{\it Left: } Inverse transfer functions $T^{-1}(L)$ for ACT DR6 and \Planck~PR4 lensing, with errors on the mean shown for each bandpower; the functions depicted here are \textit{multiplied} by the measurement bandpowers before being passed into the likelihood ($T(L)$ would be divided instead). We see differences in these two transfer functions due to misnormalization corrections in different survey footprints and consequently different overlap regions with DESI. {\it Right: } A consistency test to assess foreground contamination (see Section \ref{sec:foregrounds} for more details); we show the cross-correlation of a galaxy catalog built using the DESI HOD into the Websky simulations, with a foregrounds-only CMB map passed through the ACT DR6 baseline lensing reconstruction pipeline. Each redshift bin's cross-correlation with the foreground map is shown as a ratio to their respective 1$\sigma$ level as expressed in the covariance matrix.} 
    \label{fig:mc_norm_all}
    \label{fig:websky_fg}
\end{figure}

We define the transfer function as the following:
\begin{equation}
    \label{eqn:transfer-function}
    T(L) = \dfrac{\frac{1}{N} \displaystyle\sum_i^N C_{L, i}^{\hat{\kappa} X}}{C_{L,\, \rm theory}^{\kappa X}}
\end{equation}
where $C_{L, \,\rm theory}$ refers to a fiducial binned theory spectrum, $X \in \{\kappa, g\}$, and $N$ is the number of simulations. The transfer function is computed by calculating the mean cross spectra over a set of correlated simulations, in which a full-sky Gaussian realization of the lensing input potential or convergence is paired with its respective masked lensing reconstruction simulation that aims to emulate the final lensing data product. If $X = g$, the input lensing potentials or convergence maps are used to generate correlated Gaussian fields as described in Section \ref{sec:simulations} to be cross-correlated with the reconstructed lensing simulations; if $X = \kappa$, we simply cross-correlate the reconstructed lensing convergence with the input lensing convergence or potential. Simulation suites from \Planck~PR4 and ACT DR6 have been used for this analysis, and a pipeline is utilized to compute the cross-spectra over these simulation suites with proper mode-coupling treatment using \verb|NaMaster|. We proceed to use the transfer function with $X = \kappa$ after checking that it is consistent with the $X = g$ transfer function over all analysis scales; this choice is motivated by the $X = g$ result being noisier with greater uncertainties without using additional iterations with galaxy simulations. The inverse of the transfer functions computed for both \Planck~and ACT are shown in Figure \ref{fig:websky_fg}.

Once computed, we simply divide our cross-correlation measurement by our transfer function, ensuring that the transfer function is binned with the exact same scheme as the data bandpowers of the galaxy-CMB lensing cross power spectrum. We note that in the companion paper  \cite{Sailer_2024}, the transfer function is referred to as the ``Monte Carlo (MC) norm correction'' that is calibrated using a slightly different approach.  That approach does the following: (1) it re-applies the mask to each of the maps whenever a cross-correlation is calculated (both for the data bandpowers as well as the simulations used in the transfer correction) leading to slightly different bandpowers as well as a correspondingly different transfer function used to calculate this correction, and (2) the numerator of Equation \ref{eqn:transfer-function} is replaced with an $L$-by-$L$ power spectrum calculation of the input lensing convergence auto-spectrum using the galaxy and CMB lensing masks. Differences between the approaches can be found due to the effect of remasking a map without using a proper subset of the previously applied mask (as is the case for the ACT DR6 lensing products) as well as the uncertainty in not using \verb|NaMaster| to recover the fiducial theory spectrum $C_L^{\kappa \kappa}$ used to generate the input simulations.  However, across all analysis multipoles, we find agreement to $<0.2 \sigma$ of the inferred lensing amplitude ($A_{\rm lens}$, see equation \ref{eqn:alens}) fit to each method's corrected $C_L^{\kappa g}$ bandpowers for each of the redshift bins (with $<0.1 \sigma$ agreement for all four redshift bins jointly fit). These negligible differences are expected because of the slightly different effective masks in the two methods, which leads to slightly different areas over which the cross-correlation is measured. In Appendix C of \cite{Sailer_2024}, an explicit comparison of cosmological constraints using these two methods is presented, showing excellent agreement to well within $<0.1 \sigma$.
  
\subsection{Covariance matrix}
\label{sec:covmat}
To incorporate all of the covariance information between our cross-correlation measurements and galaxy auto-spectrum measurements, we construct a data vector:
\begin{equation*}
    \left[\{C_L^{\kappa g_i}, C_L^{g_i g_i} \,\,| \,\, \forall i \in \{1,2,3,4\}\} \right]
\end{equation*} and its respective covariance matrix:
\begin{equation*}
    \text{Cov}\left(C_L^{AB}, C_{L'}^{CD}\right)
\end{equation*}
for $\{AB, CD\} \in \{\kappa g_i, g_j g_j\}$ and $i,j \in \{1,2,3,4\}$, where the indices represent the various redshift bins.

We first build a simulation-based covariance matrix from the 400 Gaussian simulations of the CMB that are passed into the ACT DR6 lensing reconstruction pipeline. However, to reduce the noise in the estimated matrix, we draw 10 Gaussian galaxy simulations using Equations \ref{eqn:sim-gen} and \ref{eqn:sim-gen-uncorr} for each of the 400 lensing convergence simulations, yielding a total set of 4000 galaxy-CMB lensing cross-spectrum bandpowers solely generated from simulations. The final simulation-based covariance matrix is computed by the element-by-element covariance between our set of 4000 simulation cross-spectrum bandpowers, and is computed independently for each galaxy redshift bin. 

The above procedure gives a good estimate of the main diagonal of the covariance matrix, but does not capture correlations between various redshift bins. We choose not to generate and utilize ``intra-correlated'' galaxy simulations (within different redshift bins) due to the computational effort required to estimate covariances using $\mathcal{O}(10^5)$ mode-decoupling iterations for an ultimately subdominant region of our analysis covariance matrix. Instead, to capture these correlations, we build an analytic Gaussian covariance matrix (using the \verb|gaussian_covariance| method from \verb|NaMaster| \cite{García-García_2019}). This is built from pairs of angular power spectra of multiple Gaussian masked fields, by doing the following:
\begin{itemize}
    \item Taking in as input a set of fiducial theory spectra for $C_L^{\kappa \kappa}$, $C_L^{\kappa g_i}$, and $C_L^{g_i g_j}$ where $i, j$ span all galaxy redshift bin combinations.
    \item Taking in as input the effective reconstruction noise curves for the lensing measurement $N_L^{\kappa \kappa}$ as well as a fiducial galaxy shot noise spectrum $N_L^{g_i g_i}$.
    \item Computing the following\footnote{This approximation, as detailed in \cite{García-García_2019, Efstathiou_2004}, is valid if the diagonal of the coupling matrix is dominant which is true for our analysis.}:
    \begin{align*}
        \text{Cov}\left(C_L^{AB}, C_{L'}^{CD} \right) &\approx C_{(L}^{AC} C_{L')}^{BD} M_{L L'} (m_A m_C, m_B m_D) \\
        &+ C_{(L}^{AD} C_{L')}^{BC} M_{L L'} (m_A m_D, m_B m_C)
    \end{align*}
    where $C_{(L} D_{L')} =\left(C_L D_{L'} + C_{L'} D_{L} \right) /\, 2$ and the mode-coupling matrix $M_{L L'}$ is computed as a function of the mask $m_X$ of field $X$. For our purposes of pseudo-$C_L$ bandpower covariances, this is the bandpower-windowed and mode-coupled version of the expression when Wick's theorem for four fields is applied to Equation \ref{eqn:power-spectrum}. 
\end{itemize}
At the level of precision assumed for the covariance matrix, these steps result in good approximations to the true signal and noise components of the relevant power spectra. The fiducial theory spectra used for covariance estimation incorporates the same theory lensing auto-spectrum $C_L^{\kappa \kappa}$ as the one used to generate the ACT DR6 lensing reconstruction simulations used in \cite{farren2023atacama} and \cite{qu2023atacama}, but also uses theory power spectra predictions best-fit to measurements (using the \Planck~PR4 lensing convergence map) for the galaxy-CMB lensing cross-spectra $C_L^{\kappa g_i}$ for each galaxy redshift bin $i$ as well as the galaxy-galaxy power spectra $C_L^{g_i g_j}$ (see Section 3 of the companion paper \cite{Sailer_2024} for further details). We ensure that our blinding policy (Section \ref{sec:blinding}) is upheld by fitting to an already unblinded measurement while fixing our assumed cosmology.

Our final covariance matrix is a hybrid combination of the simulation-based matrix and the analytic covariance matrix: while the analytic covariance matrix provides a prediction for $\text{Cov}\left(C_L^{\kappa g_i}, C_{L'}^{\kappa g_j}\right)$, $\text{Cov}\left(C_L^{g_i g_i}, C_{L'}^{g_j g_j}\right)$, and $\text{Cov}\left(C_L^{\kappa g_i}, C_{L'}^{g_j g_j}\right)$, the simulation-based covariance matrix predicts the first two for only the case where $i = j$ (the ``on-diagonal" terms) while potentially capturing non-idealities in the CMB lensing reconstruction noise and higher-order correlations with large-scale structure. We first ensure that the analytical covariance agrees up to $\leq 5\%$ with a simulation-based covariance for the ACT DR6 $\times$ DESI and \Planck~PR4 $\times$ DESI cross-spectrum diagonals. Then, we scale the values in the analytic matrix by a multiplicative factor such that the diagonal matches that in the simulation-based matrix but making sure that the correlation coefficients are the same as that of the analytic matrix, using the following relation:
\begin{equation}
\label{eqn:hybrid-covmat}
    \mathcal{C}_{ij}^{\rm hybrid} = \mathcal{C}_{ij}^{\rm theory} \,\sqrt{\dfrac{\mathcal{C}_{ii}^{\rm sims} \mathcal{C}_{jj}^{\rm sims}}{\mathcal{C}_{ii}^{\rm theory} \mathcal{C}_{jj}^{\rm theory}}}
\end{equation}\\
where $\mathcal{C}$ is the full covariance matrix $\text{Cov}\left(C_L^{AB}, C_{L'}^{CD}\right)$. 

\begin{figure*}
\includegraphics[width=0.5\columnwidth]{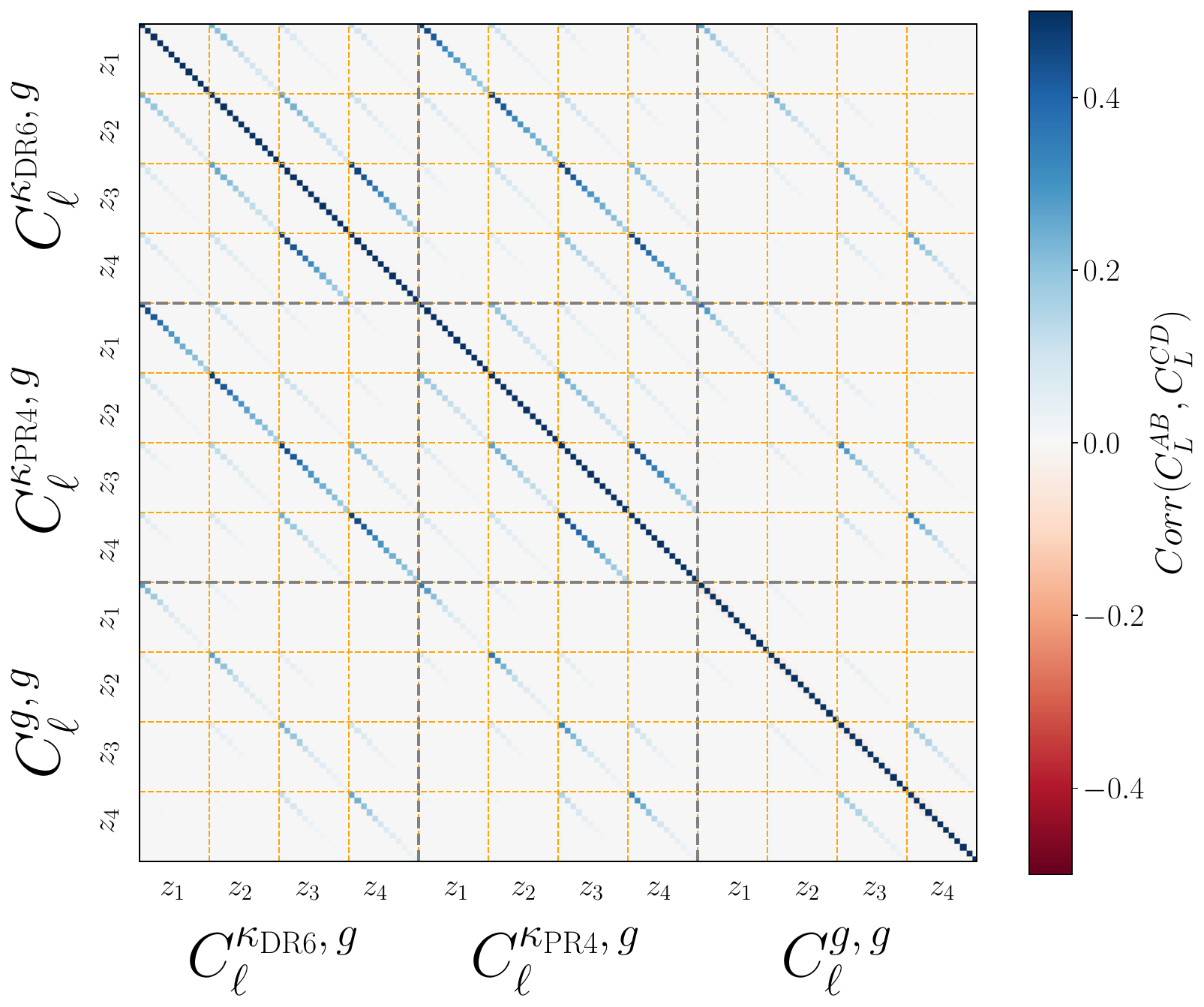}
\includegraphics[width=0.5\columnwidth]{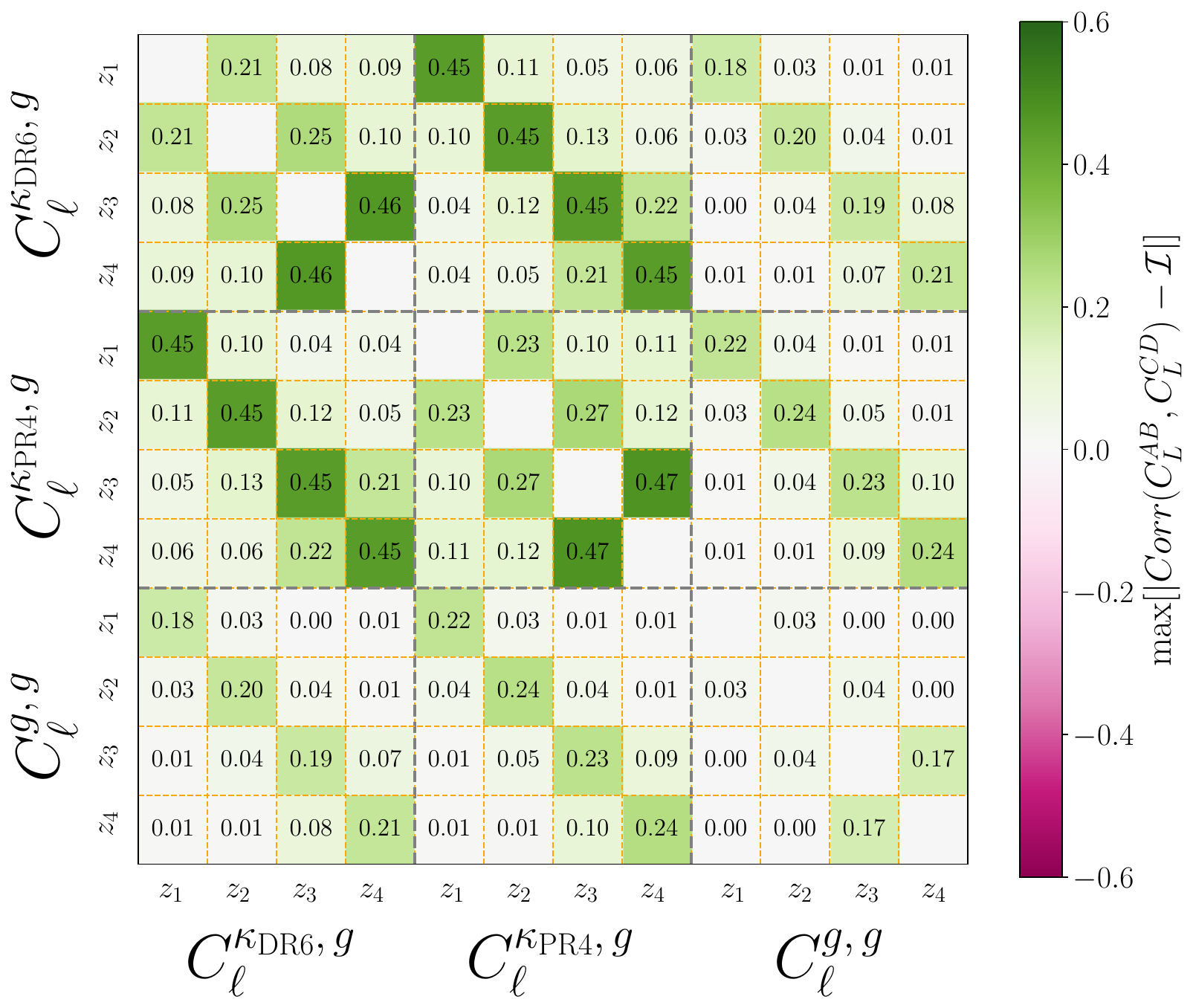}
  \caption{{\it Left: } ACT DR6 + \Planck~PR4 joint correlation matrix with the galaxy auto-spectrum from the DESI LRGs included, built using the hybrid covariance matrix described in Section \ref{sec:covmat}. Each small square represents a bandpower, ranging from $L = 20$ to $1000$. {\it Right: } Same as left, but showing the maximum correlation of each $\text{Cov}\left(C_L^{AB}, C_{L'}^{CD}\right)$ sub-block instead; the maximum correlation is computed over an analysis $L$ range common to the specific combination of spectra. The main diagonal of the full correlation matrix is removed for visual purposes here.}
    \label{fig:joint_covmat}
\end{figure*}

To assess the reliability of our estimate of the covariance matrix, we do the following: first, we compare the diagonal of the simulation-based and analytic covariance matrices; and second, a chi-squared $\chi^2 = {\bf d}^TC^{-1}{\bf d}$ computation of the measured data bandpowers ${\bf d}$ using the analytic covariance matrix described above as well as the simulation-based covariance matrix that uses varying numbers of realizations to compute the covariance. In our comparisons, using 10 Gaussian draws for the simulated galaxy fields for each of the 400 / 480 (for ACT / \Planck~respectively)  CMB lensing simulations results in values of the chi-squared metric that are consistent with the 1 Gaussian draw case to approximately 3\%. For our purposes, we do not need to include the Hartlap factor \cite{Hartlap_2007} as the correlation coefficients of the hybrid covariance matrix are all computed without simulation iterations.

For cosmology runs where we combine ACT and \Planck~lensing, we construct a joint covariance matrix. We use the data vector:
\begin{equation*}
    \left[\{C_L^{\kappa g_i} (\text{DR6}), C_L^{\kappa g_i} (\text{PR4}), C_L^{g_i g_i}  \,\,| \,\,\forall i \in \{1, 2, 3, 4\} \} \right]
\end{equation*} to construct its covariance matrix:
\begin{equation*}
    \text{Cov}\left(C_L^{AB}, C_{L'}^{CD}\right)
\end{equation*}
this time for $\{AB, CD\} \in \{\kappa_{\rm DR6} \, g_i, \kappa_{\rm PR4} \, g_i, g_i g_j\}$ and $\forall i,j \in \{1,2,3,4\}$.\\
    
For each block $\text{Cov}\left(C_L^{AB}, C_{L'}^{CD}\right)$, the analytic covariance matrix is computed as described above. If $AB = CD$ (an auto-covariance block), we have a simulation-based covariance computation to which we scale our analytic covariance with using equation \ref{eqn:hybrid-covmat}. One non-trivial section of this joint covariance matrix is the $\text{Cov}\left(C_L^{\kappa_{\rm DR6} g_i}, C_{L'}^{\kappa_{\rm PR4} g_j}\right)$ block, where we would need to estimate $C_L^{\kappa_{\rm PR4} \times \kappa_{\rm DR6}}$, or the lensing cross-spectrum between the ACT DR6 and \Planck~PR4 lensing convergence maps in order to provide input spectra for the analytic covariance calculation. We do this by using the corresponding sets of reconstructed lensing simulations for \Planck~and ACT in our cross-spectrum pipeline, and using the ensemble average of these in the analytic covariance calculation. Visualized in Figure \ref{fig:joint_covmat}, this results in this block of the covariance accurately capturing the at most approximately 40–50\% correlation between the \Planck~and ACT measurements, which share significant sky area. Looking at correlations between cross-spectra and galaxy auto-spectra, \Planck~PR4 and DESI see a maximum correlation of around 25\% while ACT DR6 and DESI see around 20\%. Since each block $\text{Cov}\left(C_L^{AB}, C_{L'}^{CD}\right)$ is of size $12 \times 12$ (with entries for each bandpower between $L = 20$ and $1000$), the full analysis covariance matrix has dimensions $144 \times 144$.

\section{Systematics and null tests}
\label{sec:nulls}
We describe here a suite of tests we have performed to ensure that the ACT cross-correlation bandpower results used in our analysis are robust. We refer the reader to \cite{Sailer_2024} for the corresponding tests for the auto-spectrum of DESI LRGs. 

\subsection{Foreground contamination assessment}
\label{sec:foregrounds}
CMB lensing maps are reconstructed from millimeter-wavelength observations (primarily at 90 and 150 GHz) that contain additional signals including the tSZ and kSZ effect, the CIB, radio sources and Galactic foregrounds. Since CMB lensing derives information significantly from higher multipoles $\ell > 2000$ of the millimeter-wavelength maps, extragalactic foregrounds adding small-scale fluctuations are the main possible source of contamination, particularly for high-resolution experiments like ACT. Many algorithmic improvements on the standard quadratic estimator have been proposed and adopted to mitigate contamination, including multi-frequency methods \cite{MMHill,Sailer:2021vpm,Darwish:2021ycf} and geometric methods \cite{Namikawa_2013,Osborne_2014,Schaan:2018tup,Sailer2020ProfileHardening,Sailer:2022jwt}.

Our baseline analysis uses a tSZ profile hardened estimator \cite{Sailer2020ProfileHardening} to mitigate foreground contamination. While this has been shown to be effective for the ACT DR6 CMB lensing auto-spectrum in \cite{maccrann2023atacama} and various tests for the unWISE cross-correlation analysis in \cite{farren2023atacama}, here, we extend that analysis to specifically assess any contamination in a cross-correlation of the lensing map with DESI LRGs. 

We create mock LRG maps from the Websky \cite{Stein:2020its,Li:2021ial} halo catalogs as follows. We weight the Websky halos by a stochastic factor $N_\text{cent}+N_\text{sat}$, where the number of centrals ($N_\text{cent}=0$ or 1) is drawn from a binomial distribution with mean $\overline{N}_\text{cent}$ and the number of satellite galaxies $N_\text{sat}$ is drawn from a Poisson distribution with mean $\overline{N}_\text{sat}$. The values of $\overline{N}_\text{cent}$ and $\overline{N}_\text{sat}$ are determined as a function of halo mass following a halo-occupation distribution (HOD) as described in \cite{Zheng_2007} (see e.g., Equations 4 \& 5 of \cite{Yuan:2023ezi}) with parameters\footnote{Specifically, we use the best fit values listed in the \cite{Zheng_2007} + $f_\text{ic}$ column of Table 3 \cite{Yuan:2023ezi}, with the exception of $f_\text{ic}$ which we set to $1$, and the cutoff mass $M_{\rm cut}$ which is tuned to match the measured large-scale clustering (at $\ell \simeq 100$) of the LRGs.} obtained from a recent fit to the DESI 1\% survey LRGs \cite{Yuan:2023ezi}. For each redshift bin, we then randomly downsample the weighted halos (by a factor of $0.4-0.55$) to match the measured shot noise of the LRG samples and reweight the remaining halos by their spectroscopically calibrated redshift distributions. We finally bin the weighted halos into \verb|HEALPix| pixels with \verb|Nside = 2048|. The power spectra of the mock LRGs differ from the data by at most 15\% on the scales relevant for our analysis, which is not a concern as these mocks are only used to qualitatively assess foreground contamination and not used to calibrate data products or theory modeling (following the reasoning presented in \cite{Krolewski_2021}).

We then cross-correlate these mock LRG maps for each redshift bin with a map that was prepared in \cite{maccrann2023atacama} by including the tSZ, kSZ and CIB signals but excluding the lensed CMB; this map is the result of the co-adding and subsequent bias-hardened reconstruction pipeline run on the Websky ``foregrounds-only" temperature field. This reconstruction uses the temperature-only quadratic estimator as we assume correlations of extragalactic foregrounds with CMB polarization are highly subdominant. Since the quadratic estimator reconstruction is heuristically a 2-point function in the CMB temperature field $\langle T T \rangle$, the cross-correlation with DESI LRGs is only biased through bispectra of the form $\langle T_f T_f \delta_g \rangle$, where $T_f$ is a foreground contaminant and $\delta_g$ is the DESI LRG overdensity: this means including the lensed CMB would only add noise and not inform our estimation of the bias.  As demonstrated in Figure \ref{fig:websky_fg}, we find that the cross-correlation of Websky foregrounds with the mock LRGs is consistent with null within our error bars. We note that since our baseline map also includes polarization data and our errors are estimated from the fiducial minimum-variance (MV) reconstruction, it is even more robust than what is suggested by this analysis. 

We quantify the consistency of the foreground bias with null through the amplitude bias parameter $\Delta A_{\rm lens}$; this is defined as a change in the amplitude of the baseline power spectrum measurement due to the contribution from the foreground-only cross-spectrum $C_{L, fg}^{\kappa g}$ (estimated as described above) relative to our fiducial galaxy-CMB lensing cross-spectrum measurement $C_L^{\kappa g}$. Following \cite{maccrann2023atacama}, we have for the amplitude bias and its uncertainty:
\begin{align}
\label{eqn:alens}
    \Delta A_{lens} &= \dfrac{\displaystyle\sum_{L L'} \left(C_{L, fg}^{\kappa g}\right)^T \text{Cov}^{-1}_{L L'} \, C_{L'}^{\kappa g}}{\displaystyle\sum_{L L'} \left(C_{L}^{\kappa g}\right)^T \text{Cov}^{-1}_{L L'} \, C_{L'}^{\kappa g}}, \,\,\,\,  \sigma_{A_{lens}} = \dfrac{1}{\sqrt{\displaystyle\sum_{L L'} \left(C_{L}^{\kappa g}\right)^T \text{Cov}^{-1}_{L L'} \, C_{L'}^{\kappa g}}} \\
    \Delta A_{lens} \, / \, \sigma_{A_{lens}} &= \dfrac{\displaystyle\sum_{L L'} \left(C_{L, fg}^{\kappa g}\right)^T \text{Cov}^{-1}_{L L'} \, C_{L'}^{\kappa g}}{\sqrt{\displaystyle\sum_{L L'} \left(C_{L}^{\kappa g}\right)^T \text{Cov}^{-1}_{L L'} \, C_{L'}^{\kappa g}}}.
\end{align}
The $\Delta A_{lens}$ for the cross-correlations of the foreground-only Websky realization with each of the four redshift bins is shown in Figure \ref{fig:websky_fg}. As all of the values of the amplitude shifts are on the order of $0.1 \sigma$ or lower, we safely assume that our galaxy sample is not significantly contaminated by foregrounds such as the tSZ, CIB, and point sources.  We will next see that apart from this simulation-based assessment, several empirical null and consistency tests performed below add further confidence to the robustness of our measurement. 

\subsection{Null tests}
\label{sec:null-tests}
We have performed a suite of null tests to ensure that our baseline galaxy-CMB lensing cross-correlation measurement is not contaminated by systematics such as biases from extragalactic foregrounds and instrumental systematics. The analyses in \cite{qu2023atacama,maccrann2023atacama} demonstrate that the ACT DR6 lensing map is robust at the level of the CMB lensing auto-spectrum, but does not eliminate the possibility of bispectrum biases (in the auto-spectrum as well as cross-correlations with large-scale structure) and Galactic contaminants correlated with residual systematics in our LRG sample (e.g., stars or extinction).

\begin{figure*}
\includegraphics[width=\textwidth]{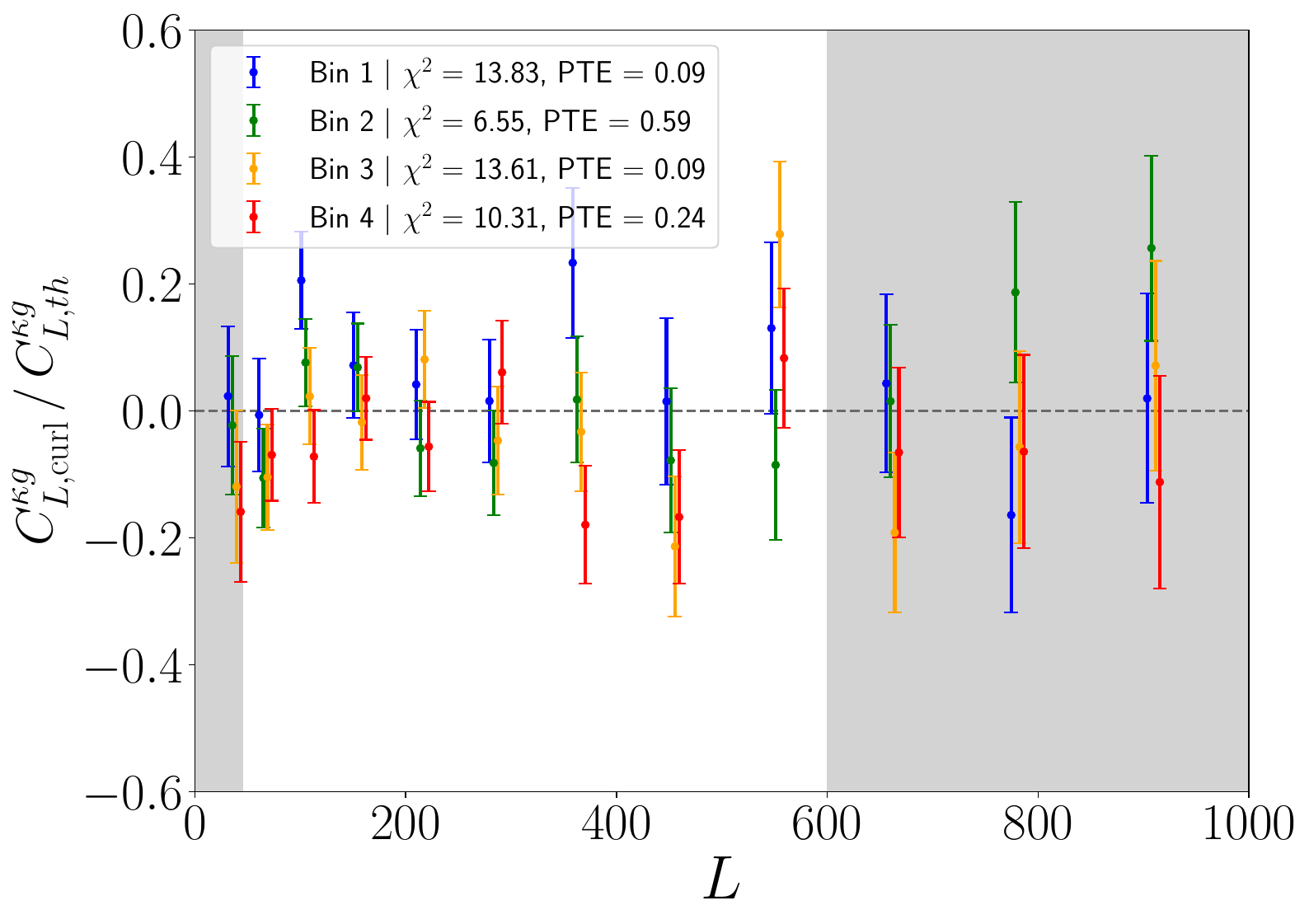}
\caption{Curl null test as described in Section \ref{sec:map-level-null}, where the curl component of the lensing convergence field is cross-correlated with the four redshift bins of our galaxy sample and depicted here as quotients with the theory predictions of cross-correlation power spectra. All four null tests computed in the analysis $L$ range pass by having PTE values between 0.05 and 0.95 demonstrating that all tests are statistically consistent with a null result.}
\label{fig:null-curl}
\end{figure*}

Our null tests are designed as $\chi^2$ tests, with a null spectrum being the assumed null hypothesis and our rejection criterion set to be a two-sided 10\% confidence level, leading to an expected 10\% uncorrelated failure rate over all tests due to statistical fluctuations. The probability-to-exceed (PTE) the obtained $\chi^2$ is then, in terms of its cumulative distribution function (CDF):
\begin{equation}
\text{PTE} = 1 - \text{CDF}_{\chi^2} (\chi^2 / n_{dof})
\end{equation}
where $n_{dof}$ refers to the number of degrees of freedom of the $\chi^2$ computation, equal to the number of bandpowers in our null spectrum. The $\chi^2$ is computed as the following:
\begin{equation}
\chi^2 = \textbf{d}^T_{L} \text{Cov}^{-1}_{L L'} \textbf{d}_{L'}
\end{equation}
for our null data bandpower vector $\textbf{d}$ and its covariance matrix, computed over the analysis $L$ range as defined in Section \ref{sec:angular-power-spectrum}. By construction, failures can be defined and caused by two outcomes: a $\chi^2$ value large enough to result in a PTE $< 0.05$ allows us to reject the null hypothesis and conclude that a non-null signal is statistically significant, while a $\chi^2$ value small enough to result in a PTE $> 0.95$ tells us that either our computed bandpowers $d$ agrees with the null spectrum better than statistically expected, or that our covariance overestimates the error levels for $d$. While Section \ref{sec:covmat} described how the hybrid covariance matrix for our baseline cosmology data vector is constructed from theory and simulations, here, for null tests, we use different covariance matrices constructed entirely from simulations following the decision of previous analyses using these lensing products such as \cite{farren2023atacama}. To correct the inverse of our simulation-based covariance matrix appropriately, we make sure to apply the Hartlap correction factor from \cite{Hartlap_2007}:
\begin{equation}
    \text{Cov}^{-1}_{\rm corr.} = \dfrac{n - p - 2}{n - 1} \times \text{Cov}^{-1} 
\end{equation}
where $n$ is the number of data samples used to estimate the covariance of a $p$-sized data vector. As the ACT DR6 lensing suite contains 400 CMB simulations and the analysis $L$ range described in Section \ref{sec:measurement} consists of 8 bandpowers, the Hartlap correction factor affects the $\chi^2$ value by approximately 2\%.
In accordance with our baseline cross-correlation analysis, we apply the appropriate transfer functions for each of the data products, noting that some null data maps may feature different footprints and masks.

\begin{table*}[!htb]
    \centering
    \begin{tabular}{ |p{9cm}||p{1cm}|p{1cm}|p{1cm}|p{1cm}|}
        \hline
        \multicolumn{5}{|c|}{Current null test PTEs} \\
        \hline
        \textbf{Null test}& $z_1$ & $z_2$ & $z_3$ & $z_4$\\
        \hline
        $QE(\text{curl}) \times g$ & 0.086 & 0.586 & 0.093 & 0.244 \\
        $QE(\texttt{f150} - \texttt{f090} \text{ MV}) \times g$ & 0.490 & 0.852 & 0.538 & 0.864 \\
        $QE(\texttt{f150} - \texttt{f090} \text{ TT}) \times g$ & \textbf{0.971} & 0.135 & 0.296 & 0.130 \\
        $QE(\texttt{f150} \text{ MV}) \times g - QE(\texttt{f090} \text{ MV}) \times g$ & 0.631 & 0.862 & 0.891 & 0.671 \\
        $QE(\texttt{f150} \text{ TT}) \times g - QE(\texttt{f090} \text{ TT}) \times g$ & \textbf{0.995} & 0.719 & 0.945 & 0.662 \\
        $QE(\texttt{f090} \text{ MV}) \times g - QE(\texttt{f090} \text{ TT}) \times g$ & 0.325 & 0.408 & 0.583 & 0.330 \\
        $QE(\texttt{f150} \text{ MV}) \times g - QE(\texttt{f150} \text{ TT}) \times g$ & \textbf{0.971} & 0.161 & 0.263 & 0.535 \\
        $QE(\text{baseline MV}) \times g - QE(\text{baseline MVPOL}) \times g$ & \textbf{0.985} & 0.690 & 0.778 & 0.648 \\
        $QE(\text{baseline MV}) \times g - QE(\text{CIB deproj.}) \times g$ & 0.103 & 0.553 & 0.820 & 0.655 \\
        $QE(\text{baseline 60\%}) \times g - QE(\text{baseline 40\%}) \times g$ & 0.427 & 0.371 & \textbf{0.982} & 0.313 \\
        $QE(\text{baseline MV}) \times g - QE(\text{baseline MV}) \times g_{\text{DES area}}$ & 0.169 & 0.876 & 0.252 & 0.759  \\
        $QE(\text{baseline, NGC}) \times g - QE(\text{baseline, SGC}) \times g$ & 0.056 & 0.639 & 0.644 & 0.374 \\ 
        \hline
    \end{tabular}
    \caption{Here we show the results of our 48 null tests, 12 per redshift bin. Values in bold font are PTEs that lie outside of our two-sided 10\% confidence level and are treated as failures. See Section \ref{sec:null-tests} for a discussion of all of these tests, Section \ref{sec:null-test-summary} for a summary of their results, and Figures \ref{fig:null-curl}, \ref{fig:null-mvpol}, and Appendix \ref{sec:appendix-null-tests} for the plots of these tests.}
    \label{tab:null-test-results}
\end{table*}

\subsubsection{Map-level null tests}
\label{sec:map-level-null}
We compute three sets of map-level null tests, which generally involve the cross-correlation of our DESI LRG overdensity map with a null lensing reconstruction map. 

\begin{enumerate}
    \item The lensing displacement field can be decomposed into a gradient and curl component, where the former traces the lensing potential and the latter is expected to be zero at linear order. Barring post-Born corrections to lensing \cite{Pratten_2016} (that we don't expect to have sensitivity to with current data), the curl component should have a null correlation with the galaxy field. To test this, we cross-correlate the ACT DR6 curl map with our galaxy maps. As shown in Table \ref{tab:null-test-results}, all four galaxy redshift bins have a null correlation with our confidence levels, and the results are shown in Figure \ref{fig:null-curl}.

\item The other two map-level null tests involve a subtraction of CMB maps created by ACT DR6 with the two frequency bands, \texttt{f150} and \texttt{f090}. The CMB maps measured in these two bands are subtracted to remove the lensed CMB signal, and then passed through the lensing reconstruction to generate convergence maps. In addition to our baseline estimator which uses a MV combination of quadratic estimators (QEs) run on temperature and polarization data, we also perform temperature-only (TT) reconstructions. This is a powerful null test since it removes the large source of variance from the reconstruction noise arising from the primary CMB fluctuations themselves. Residuals in the map difference primarily include foregrounds such as the tSZ and CIB that have different amplitudes at 90 and 150 GHz. The QE pipeline includes our baseline profile hardening foreground mitigation, so we expect this test to pass when these null lensing maps are cross-correlated with the DESI LRG overdensity maps.

\end{enumerate}

As seen in Table \ref{tab:null-test-results}, these three map-level null tests are performed for each of the four redshift bins and generally pass, except for $QE(\texttt{f150} - \texttt{f090}\,\, \text{TT}) \times g$ (bin 1, PTE = 0.971).

\subsubsection{Bandpower-level null tests using frequency splits}
We run four sets of null tests involving CMB splits that differ from the map-level null tests in the fact that they are first individually passed through the lensing reconstruction pipeline before being subtracted at the spectrum level. As each of the cross-spectra with DESI LRGs are computed, they are corrected for their appropriate transfer function (see Section \ref{sec:transfer}) using the appropriate set of simulations designed for these specific null tests. The two \texttt{f150} and \texttt{f090} CMB maps have their lensing signal reconstructed using each of the aforementioned MV and TT estimators, and then subtracted in two ways:
\begin{itemize}
    \item \textbf{Different frequency, same QE} -- this is the bandpower-level version of the frequency split map-level null tests that ensures that there is no excess signal that is present in one CMB frequency's cross-correlation with the galaxies with respect to the other CMB frequency. 
    \item \textbf{Same frequency, different QE} -- this now checks at the bandpower level if there is excess signal present in a galaxy cross-correlation with the MV estimator compared to the TT estimator, and vice versa. 
\end{itemize}
These 16 tests lead to 14 passes and 2 failures: $QE(\texttt{f150} \,\,\text{TT}) \times g - QE(\texttt{f090} \,\,\text{TT}) \times g$ (bin 1, PTE = 0.995) and $QE(\texttt{f150} \,\,\text{MV}) \times g - QE(\texttt{f150} \,\,\text{TT}) \times g$ (bin 1 = 0.971). We have assessed whether these high PTE failures are due to mis-estimation of the covariance by comparing with an analytic version. A manipulation of the Gaussian covariance expression allows us to estimate the covariance of the spectrum-level difference as the following:
\begin{align*}
\text{Cov}\left[C_L^{\kappa_1 g} - C_L^{\kappa_2 g}, C_L^{\kappa_1 g} - C_L^{\kappa_2 g} \right] &= \text{Cov} \left[C_L^{(\Delta \kappa) g}, C_L^{(\Delta \kappa) g} \right] \,\,\,\,\,\,\{ \Delta\kappa \equiv \kappa_1 - \kappa_2\} \\
&= \dfrac{1}{\Delta L (2L+1)} \times \dfrac{1}{f_{sky}} \times \left(C_L^{\Delta \kappa \Delta \kappa} + N_L^{\Delta \kappa \Delta \kappa} \right) \times \left(C_L^{gg} + N_L^{gg} \right)
\end{align*}
where $\Delta L$ is the difference in the binned centers of two consecutive bandpowers. This result allows us to cross-check our Monte Carlo simulation-based covariance and confirm that the errors on our bandpowers are in good agreement -- we attribute these marginal failures to statistical fluctuations.

\subsubsection{Bandpower-level null tests using the baseline lensing map}
\begin{figure}
\includegraphics[width=0.59\textwidth]{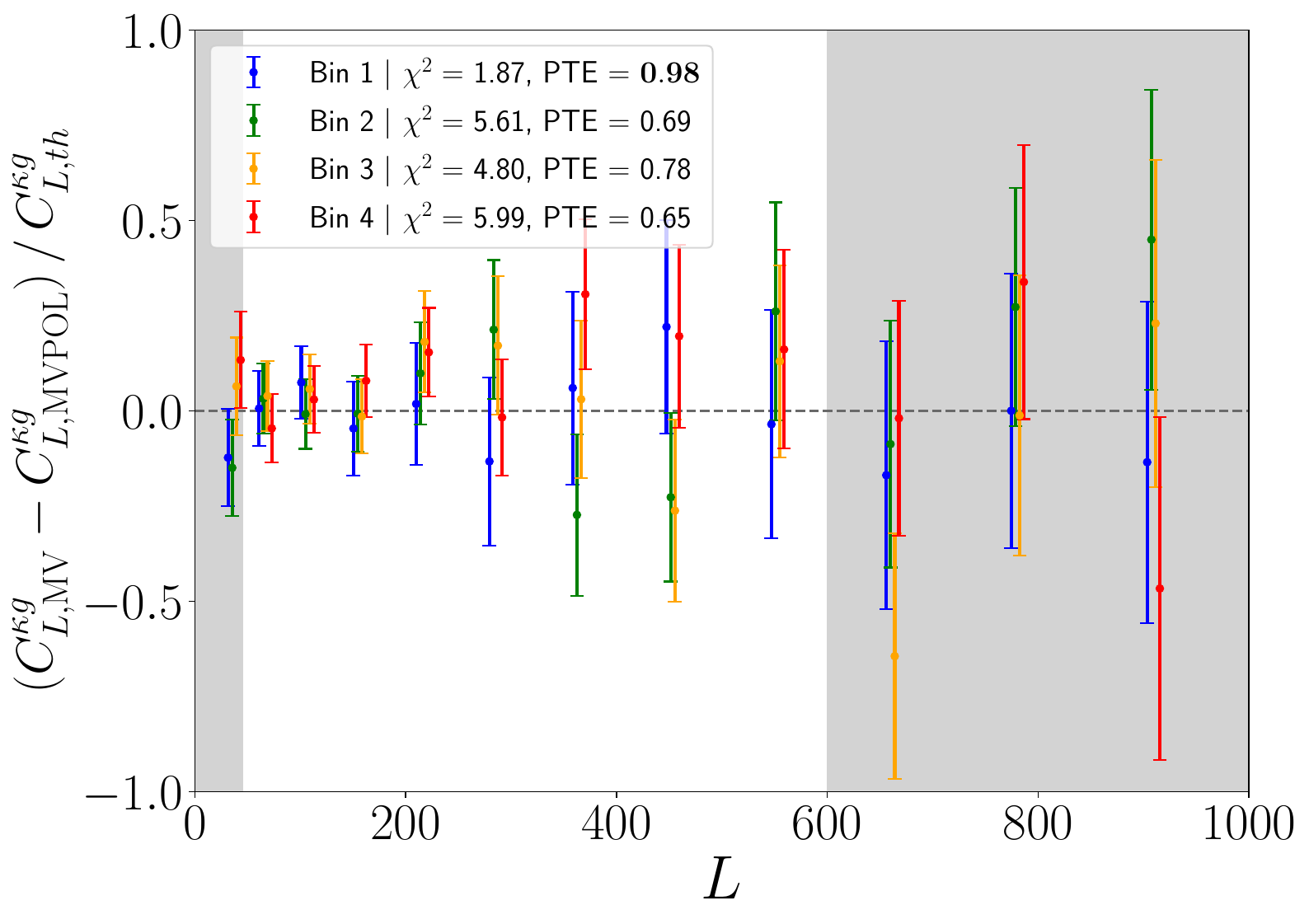}
\includegraphics[width=0.40\columnwidth]{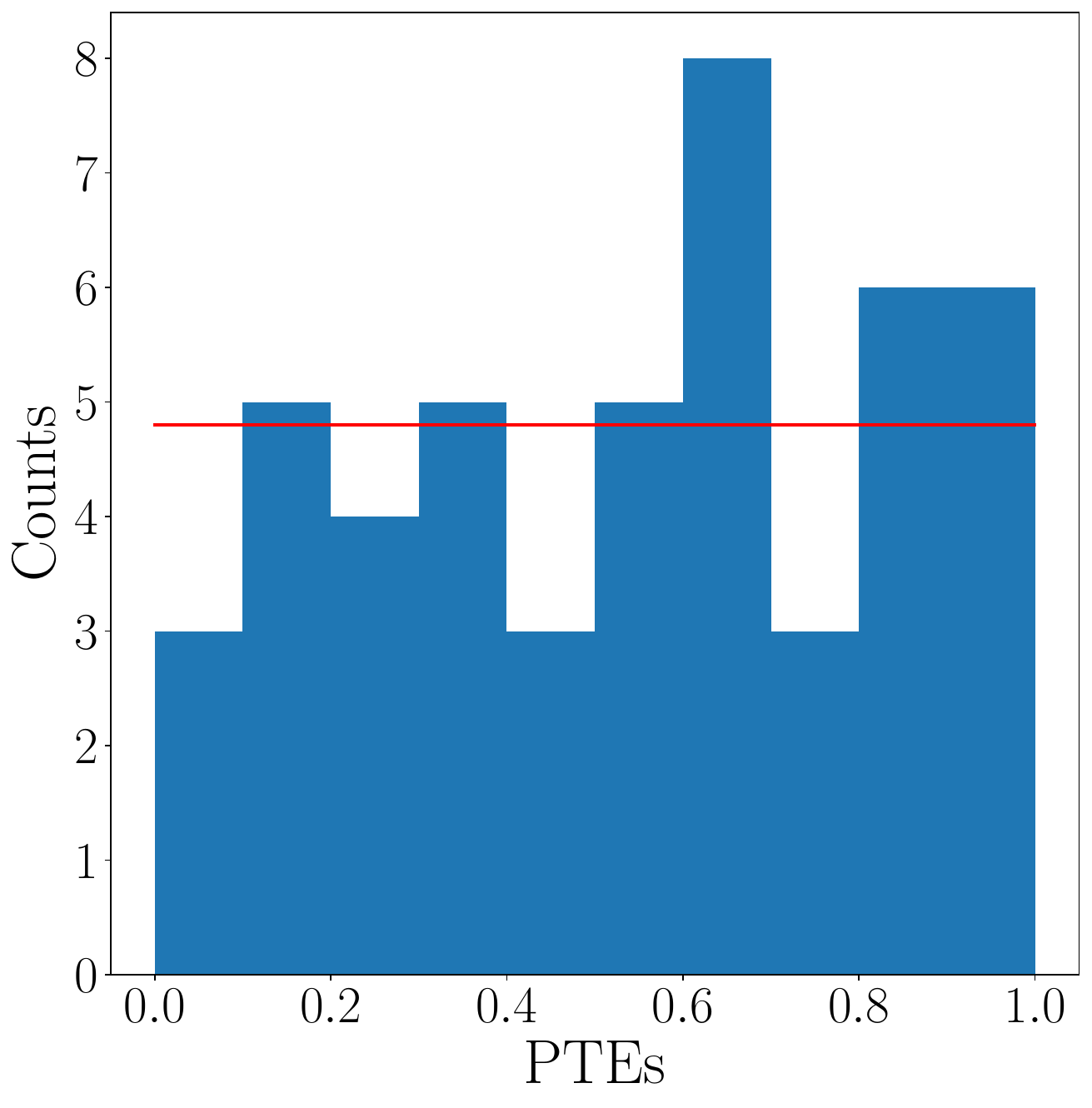}
\caption{{\it Left:} The bandpower spectrum-level null test between the baseline MV reconstruction for $C_L^{\kappa g}$ and baseline MVPOL reconstruction for $C_L^{\kappa g}$, of which $z$-bin 1 ``fails" due to an excessive PTE. {\it Right:} A histogram of the PTE distribution of the 48 null tests run for this analysis, showing its relative uniformity as well as assurance that null test failures are not systematically driven towards high or low PTE values specifically.}
\label{fig:null-mvpol}
    \label{fig:pte_histogram}
\end{figure}

The remaining null tests are now computed using the baseline MV-reconstructed lensing map, and comparing its cross-correlation with the galaxy map to those using different variants of the lensing product, by subtracting their respective cross-correlation bandpowers. This includes the following:
\begin{itemize}
    \item \textbf{Minimum variance with polarization only}. This is the lensing reconstruction run using the minimum variance polarization (MVPOL) estimator, which uses a minimum variance combination of the EE and EB quadratic estimator reconstructions. The polarization-only map is expected to be more robust against foreground contamination at the cost of significant degradation in signal-to-noise. The results of this null test are shown in Figure \ref{fig:null-mvpol}.
    \item \textbf{CIB deprojected}. This is an MV lensing reconstruction using a symmetrized quadratic estimator \cite{MMHill,Darwish_2023} in which the CIB is explicitly deprojected through a harmonic internal linear combination (ILC) run that includes higher frequency \Planck~maps \cite{maccrann2023atacama}, as an alternative to profile hardening. This specific reconstruction uses a slightly different lensing mask that removes a few extra patches with excess Galactic dust contamination. This null test checks if there may have been CIB  contamination in our baseline map and generally explores the robustness of our foreground mitigation.
    \item \textbf{Conservative lensing mask}. This is an MV lensing reconstruction run on a strict subset of the baseline lensing analysis mask (which is labeled \texttt{GAL060}) that covers approximately 40\% of the full sky (\texttt{GAL040}) and masks out additional regions with potential Galactic contamination. This null test checks if the baseline cross-correlation result is free of Galactic dust contamination.
    \item \textbf{DES footprint mask}. This is an MV lensing reconstruction run with the \texttt{GAL060} lensing analysis mask, but the cross-correlation is run with a more restrictive, subset galaxy mask that only contains the active observing footprint of DES imaging data. As described in \cite{Sailer_2024}, this null test checks if there is a systematic offset in the cross-correlation within the DES sub-region only, where the imaging data is deeper and the galaxy selection inside and outside of the sub-region can be non-trivially and systematically different.
    \item \textbf{NGC vs SGC}. This is an MV lensing reconstruction run with the \texttt{GAL060} lensing analysis mask, but the cross-correlation is run with the intersection of the Galaxy mask and masks that cover the North and South Galactic Caps (NGC, SGC). This null test checks if there is an extra signal or systematic in one of the Galactic hemispheres compared to the other.
\end{itemize}
We see 2 failures from this set of tests, $QE(\text{baseline MV}) \times g - QE(\text{baseline MVPOL}) \times g$ (bin 1, PTE = 0.985) and $QE(\text{baseline 60\%}) \times g - QE(\text{baseline 40\%}) \times g$ (bin 3, PTE = 0.982).
 
\subsubsection{Null test summary}
\label{sec:null-test-summary}
Again, we expect about 10\% of uncorrelated null tests to fail due to statistical fluctuations for our twelve sets of null tests run on all four galaxy redshift bins. Out of these total 48 runs, we report 5 total failures, shown in bold in Table \ref{tab:null-test-results}.  Based on these results, we are confident that systematics do not contribute significantly to the measurement and attribute the null test failures to statistical fluctuations, noting also the following:
\begin{itemize}
    \item Some of the null tests are correlated, either due to the usage of the same galaxy redshift bin (Bin 1 has 4 of the 5 failures) or between the different products used for different null tests ($QE(\texttt{f150} \text{ TT})$, for example, is involved in three separate failures). This also implies that the number of uncorrelated null test failures should be appropriately compared to the total number of uncorrelated null tests, which are both difficult to exactly compute. However, at face value, the failure rate of uncorrelated null tests should not significantly exceed the 10.4\% value we find with our set of 5 failures out of 48 tests.
    \item All of the null test ``failures'' are due to PTEs higher than 0.95. This suggests the following: first, these are not strictly failures in which we believe that the null test shows a statistically significant deviation from null; second, the simulation-based error levels for the null tests may be overestimated. To address this, the errors for all failures were cross-checked to be in agreement between the simulation-based covariance and an analytic Gaussian covariance. This also confirms that the non-Gaussian contributions from lensing reconstruction that are observed in the simulations but not in the analytic covariance are relatively small, and the mode-coupling effect is treated no differently when using the analytic expression or the Gaussian sims. 
    \item The distribution of PTEs is approximately uniform as expected, shown in Figure \ref{fig:pte_histogram}. This supports the idea that our PTEs are not collectively skewed towards zero or one due to a systematic across various null tests.
\end{itemize}

In addition to these null test results, \cite{Sailer_2024} confirms with parameter-level tests that by using linear theory modeling choices and scale cuts, $S_8$ is fully consistent with our baseline ACT-only constraint when using variations of the ACT DR6 lensing map such as the CIB-deprojected reconstruction, the single-frequency CMB splits, and others.

\section{Cosmological constraints and analysis}
\label{sec:cosmology}
Abiding by our blinding policy described in Section \ref{sec:blinding} and confirming that parameter-level tests (Section \ref{sec:param-recovery}) are acceptably passed, we perform a likelihood-based inference that uses a theory model (Section \ref{sec:theory}), set of priors (Section \ref{sec:priors}), and a likelihood (Section \ref{sec:inference}) to estimate a constraint on $S_8^\times$. We briefly summarize the relevant methodology in this section and leave details to the companion paper \cite{Sailer_2024}.

\subsection{Blinding policy}
\label{sec:blinding}
 To mitigate the influence of confirmation bias, we adopt a blinding policy which prohibits galaxy-CMB lensing cross-spectrum comparisons between ACT DR6 and \Planck~PR4 as well as comparisons of both results to fiducial theory. Our blinding policy consisted of two stages:
 \begin{enumerate}
     \item Blinding at the spectrum level, during which we specifically ensured that our cross-correlation bandpowers were never compared to theory predictions
     \item Blinding at the parameter level, during which we specifically ensured that we did not look at any constraints on cosmological parameters that used our unblinded cross-correlation bandpowers
 \end{enumerate}
 for which we only considered unblinding parameters after we had already unblinded our bandpowers. Before unblinding our power spectra, we ensure the following:
 \begin{itemize}
   \item The pipeline is able to reproduce results of the cross-correlation between \Planck~PR3 lensing and DESI LRGs \cite{White_2022} as well as the galaxy auto-spectrum of the DESI LRGs.
   \item The pipeline is able to recover a fiducial prediction for the galaxy-CMB lensing cross-spectrum as well as the galaxy auto-spectrum from correlated Gaussian simulations.
   \item The measurement is not contaminated significantly by Galactic and extragalactic foregrounds, tested by populating a DESI LRG-like HOD in the Websky simulations and observing a null cross-correlation signal with a foregrounds-only lensing reconstruction.
   \item The measurement is not contaminated significantly by other systematics, tested by running a null test suite across different combinations of CMB and galaxy maps and ensuring that at a two-sided 10\% rejection level of the null hypothesis, no more than the statistically expected number of null tests fail.
   \item The pipeline is able to recover input fiducial cosmological parameters using noiseless, binned theory data vectors and the analysis covariance matrix, likelihood, priors, and convergence criterion to good precision (summary in Section \ref{sec:param-recovery}, details in Section 5.4 in \cite{Sailer_2024}).
   \item The pipeline is able to recover input fiducial cosmological parameters using the \verb|Buzzard| simulations \cite{derose2019buzzard}, for which \cite{Sailer_2024} models LRG-like halos and CMB lensing to compute a noisy cross-correlation data vector (summary in Section \ref{sec:param-recovery}, details in Section 5.5 in \cite{Sailer_2024}).  
 \end{itemize}

\noindent Before unblinding our constraints, we ensure the following:
\begin{itemize}
   \item The cross-correlation measurement bandpowers between \Planck~PR4 and DESI LRGs are not statistically discrepant from the bandpowers computed for the ACT DR6 and DESI LRGs cross-correlation.
   \item The pipeline is able to then assess parameter-level consistency between blinded ACT and \Planck, HEFT and linear theory, as well as variations from our baseline analysis, including conservative scale cuts for our multipole range (see Figure \ref{fig:variations}) and additionally masking LRGs on the ACT footprint. 
\end{itemize}

 Our blinding policy did allow us to use a blinded version of the ACT DR6 lensing convergence map that contains a random multiplicative blinding factor for initial pipeline development and early iterations of some null tests; this blinded map was used in other ACT DR6 lensing analyses such as \cite{farren2023atacama} and \cite{qu2023atacama}. 
 
\subsection{Theory model}
\label{sec:theory}
We briefly summarize here our theory model, with further details found in our companion paper \cite{Sailer_2024}. We use hybrid effective field theory (\cite{10.1093/mnras/staa251}) to model predictions for theory spectra, which uses a combination of the Lagrangian perturbation theory (LPT) prediction + the Aemulus-$\nu$ simulations \cite{DeRose_2023} to model the matter density field composed of both cold dark matter and baryons. The usage of HEFT also motivates our scale cuts, as \cite{Kokron21} cites sub-percent accuracy for LRG-like halo clustering and halo-matter power spectra fitting for $k \approx 0.6 \, h \, / \, \text{Mpc}$, allowing us to probe smaller scales than what was used in \cite{White_2022}.

HEFT allows us to parameterize cosmological power spectra as a linear combination of the CDM + baryon power spectrum $P_{cb}(k)$ and various intermediate component-basis spectra $P_{i,j}(k)$ that capture two-point correlations between different overdensities expressed in the Lagrangian bias formalism. To 1-loop or second order, this linear combination is expressed using a set of Lagrangian bias parameters $b_i$ for $i = 1, 2, s$ that quantify the contribution of the CDM + baryon overdensity fields $\delta_{cb}$, $\delta_{cb}^2$, and the tidal shear field $s_{cb}$ respectively. As highlighted in \cite{Sailer_2024}, we also use counterterms $\alpha$ to capture interactions with the derivative field $\nabla^2 \delta_{cb}$ and other small-scale stochastic components. Using these bias parameters that are independently defined and varied per redshift bin along with cosmological parameters as inputs, predictions for the intermediate power spectra are computed efficiently by an emulator trained on the Aemulus-$\nu$ simulations \cite{DeRose_2023} which model, and then Limber integrated over the line-of-sight to obtain predictions for the observables $C_L^{gg}$ and $C_L^{\kappa g}$. As the theory power spectrum depends linearly on the counterterms for the galaxy auto-spectrum and the cross-spectrum with matter as well as the shot noise $SN$, we can assume a Gaussian prior for these linear parameters and analytically marginalize our likelihood with respect to them. Further details of the marginalization procedure, and its implementation in our likelihood can be seen in \cite{Sailer_2024}.

\subsection{Cosmological parameterization and priors}
\label{sec:priors}

\begin{table}
\begin{tabularx}{\columnwidth}{l @{\hskip 1.4in}l}
\hline
\hline
Parameter & Prior \\
\hline
\multicolumn{2}{l}{\bf Fixed parameters } \\
$n_s$ & 0.9649 \\
$\Omega_b h^2$ & 0.02236 \\
$\Omega_m h^3$ & 0.09633 \\
$\sum m_{\nu}$ & 0.06 {\rm eV} \\
\hline
\multicolumn{2}{l}{\bf Cosmological parameters } \\
$\log (10^{10} A_s)$ & $\mathcal{U}(2,4)$ \\
$\Omega_{c} h^2$ & $\mathcal{U}(0.08, 0.16)$\\
\hline
\multicolumn{2}{l}{\bf Analytically marginalized parameters}\\
$\alpha_a$ & $\mathcal{N}(0, 50)$ \\
$\epsilon$ & $\mathcal{N}(0, 2)$ \\
$N_L^{gg}$ & $10^{-6} \,\, \mathcal{N}(4.07 \,|\, 2.25 \,|\, 2.05 \,|\, 2.25, 0.3 \times 4.07 \,|\, 2.25 \,|\, 2.05 \,|\, 2.25)$ \\
\hline
\multicolumn{2}{l}{\bf Nuisance parameters}\\
$b_1$ & $\mathcal{U}(0, 3)$ \\
$b_2$ & $\mathcal{U}(-5, 5)$ \\
$b_s$ & $\mathcal{N}(0, 1)$ \\
$\mu_i$ & $\mathcal{N}(0.972 \,|\, 1.044 \,|\, 0.974 \,|\, 0.988, 0.1)$ \\
\hline 
\end{tabularx}
\caption{Parameters and priors used in this work and \cite{Sailer_2024}. Uniform priors from $x_1$ to $x_2$ are denoted with $\mathcal{U}(x_1, x_2)$ and Gaussian priors with mean $\mu$ and standard deviation $\sigma$ are shown as $\mathcal{N}(\mu,\sigma)$. Nuisance parameters $b_1, b_2, b_s$ are all bias parameters for the HEFT theory model; counterterms are represented with $\alpha_a$ and $\epsilon$; and $\mu_{i}$ is the magnification bias for galaxy redshift bin $z_i$. Only the shot noise spectrum $N_L^{gg}$ and magnification bias $\mu_i$ have redshift bin-dependent priors, with $\mu$ and $\sigma$ shown respectively for bins 1, 2, 3, and 4.}
\label{tab:priors}
\end{table}

We show our priors and parameterization in Table \ref{tab:priors}. To constrain the amplitude of structure, we sample over $\log (10^{10} A_s)$ and $\Omega_{c} h^2$. We fix $n_s$ and $\Omega_b h^2$ to a value preferred by \Planck~CMB measurements, the sum of neutrino masses $\sum m_{\nu}$ to the minimal value allowed by neutrino oscillation experiments and $\Omega_m h^3$ to a value informed by the precisely measured angular size of the sound horizon from \Planck~CMB measurements.

For the HEFT model, we put priors on analytically marginalized parameters ($\alpha_a$ for the auto counterterm, $\epsilon$ as a parameterization of the cross counterterm, and $N_L^{gg}$ for the shot noise), the Lagrangian bias parameters $b_1$, $b_2$, and $b_s$ (up to second or 1-loop order), and the magnification bias $\mu$. We put relatively uninformative priors on all of these HEFT parameters except for $b_s$ and $\epsilon$, where the former is found to share a strong degeneracy with $b_2$ and the latter is chosen to appropriately represent the size of small-scale effects we expect from baryonic feedback \cite{2206.11794} and our usage of the Aemulus-$\nu$ simulations. These are discussed in further detail in \cite{Sailer_2024}.

\begin{figure}
\centering
\includegraphics[width=0.495\columnwidth]{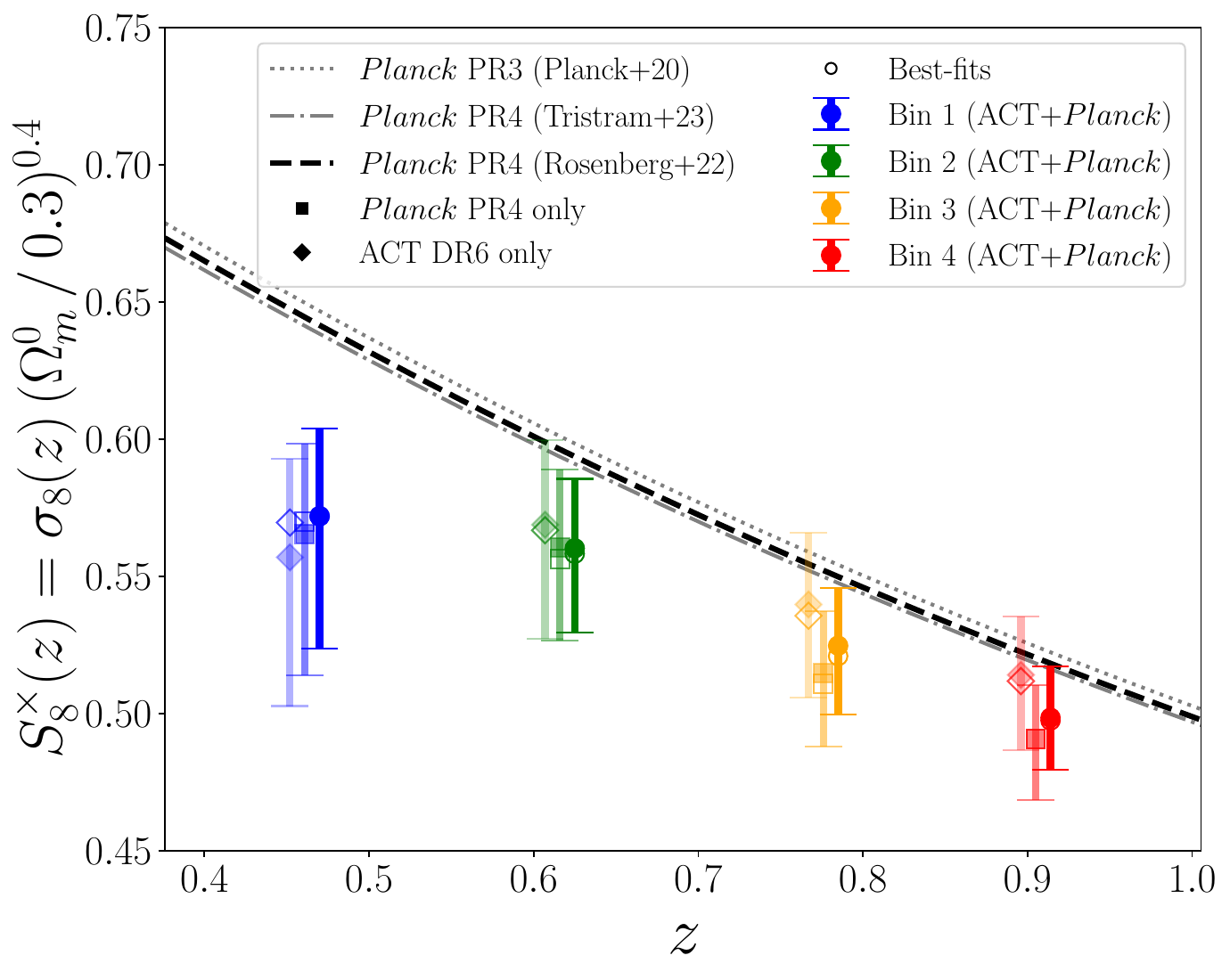}
\includegraphics[width=0.495\columnwidth]{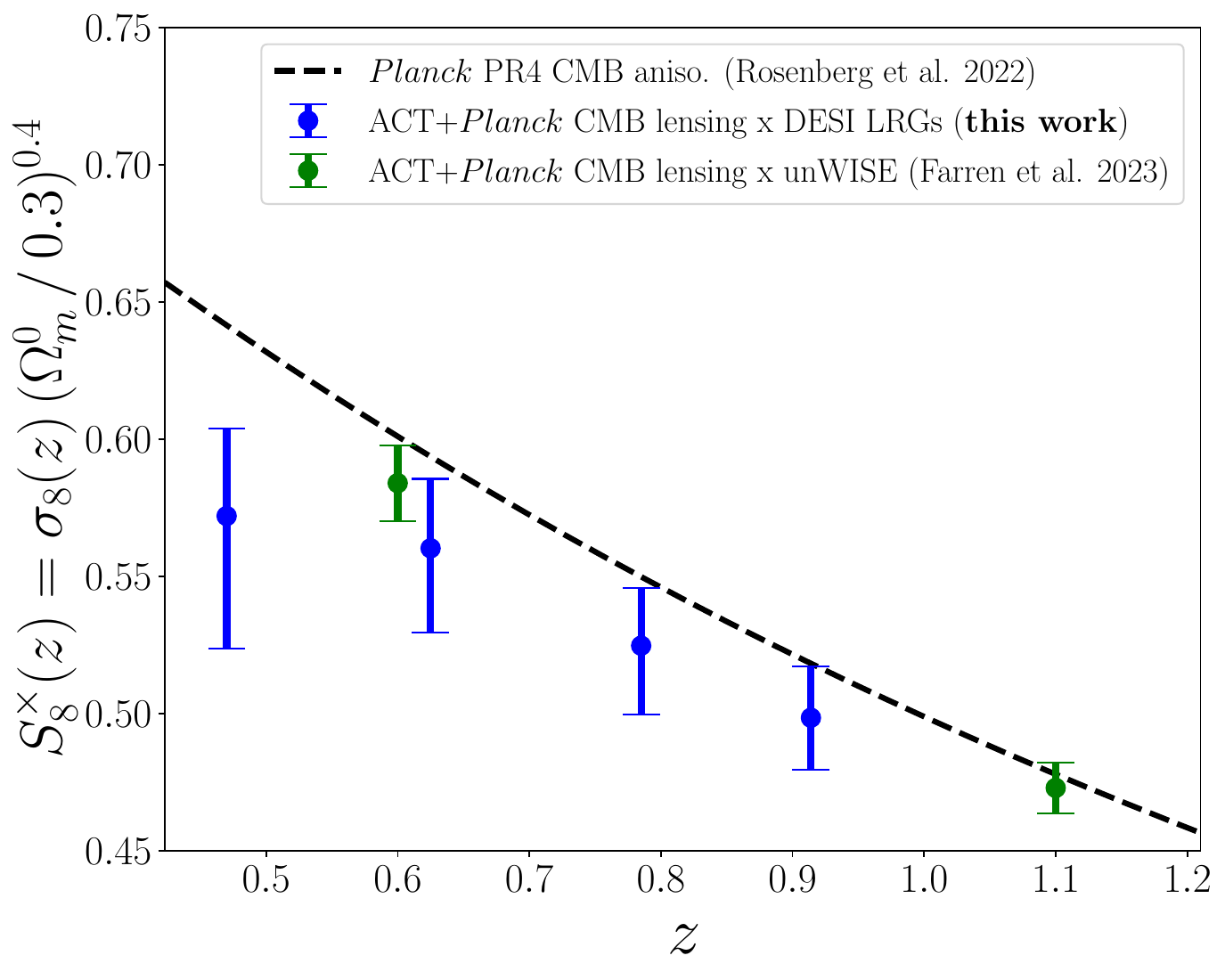}
\caption{{\it Left:} $S_8^\times(z)$ shown for our four redshift bins with the \Planck~PR4 measurement, ACT DR6 measurement, and the joint ACT + \Planck~fit. We note the means are consistent with the best-fit points shown as open markers, and that these means are consistently low compared to the \Planck~PR4 CMB prediction in concordance with the baseline joint-redshift constraint. The theory predictions computed using \texttt{CAMB} (\cite{Lewis_1999bs}, \cite{Lewis_2002ah}, \cite{Howlett_2012mh}) with the \Planck~cosmological parameters are shown in dashed lines. {\it Right:} Showing same joint ACT + \Planck~fits as left, but we also plot the unWISE cross-correlation result with \Planck~PR4, ACT DR6, and their joint fit from \cite{farren2023atacama}, which sees better agreement with the \Planck~PR3 CMB at lower redshifts.}
\label{fig:s8xz}
\end{figure}
\subsection{Parameter inference}
\label{sec:inference}
We adopt a Gaussian likelihood, taking the form:
\begin{equation}
\label{eqn:loglike}
-2 \ln \mathcal{L} \propto 
\begin{bmatrix}
\hat{C}_L^{\kappa g} - C_L^{\kappa g}(\bm{\theta})  \\
\hat{C}_L^{gg} - C_L^{gg}(\bm{\theta}) \\
\end{bmatrix}^{T}
\begin{bmatrix}
\text{Cov}(C_L^{\kappa g}, C_{L'}^{\kappa g}) & \text{Cov}(C_L^{\kappa g}, C_{L'}^{g g}) \\
\text{Cov}(C_L^{gg}, C_{L'}^{\kappa g}) & \text{Cov}(C_L^{gg}, C_{L'}^{g g}) \\
\end{bmatrix}^{-1}
\begin{bmatrix}
\hat{C}_{L'}^{\kappa g} - C_{L'}^{\kappa g}(\bm{\theta})  \\
\hat{C}_{L'}^{gg} - C_{L'}^{gg}(\bm{\theta}) \\
\end{bmatrix}
\end{equation}
where $\hat{C}_L^{AB}$ for $AB \in \{\kappa g, gg\}$ represents a power spectrum measurement, $C_L^{AB}(\bm{\theta})$ represents the prediction from the HEFT matter and galaxy power spectra using cosmological parameters $\bm{\theta}$, and the covariance blocks $\text{Cov}\left(C_L^{AB}, C_{L'}^{CD}\right)$ are computed as described in Section \ref{sec:covmat}.

The cosmological parameter space was sampled using the Markov Chain Monte Carlo (MCMC) method  with the \verb|Cobaya| framework \cite{2019ascl.soft10019T,Torrado_2021}, and best-fit values were obtained by using the \verb|minimize| sampler built in \verb|Cobaya|. The MCMC chains were sampled using the likelihood from Section \ref{eqn:loglike} until a Gelman-Rubin convergence criterion \cite{10.1214/ss/1177011136} of $R-1 < 0.01$ was reached. The first 30\% of the chains are removed as burn-in chains before the contours are visualized and analyzed using \verb|GetDist| \cite{Lewis:2019xzd}. 

The MCMC sampling is also run on each redshift bin independently, where only the nuisance parameters for each bin is sampled along with the appropriate cosmological parameters. This information from the best-fit cosmologies can be used to understand the redshift dependence of structure growth, as we can compute and plot a redshift-dependent $S_8^\times(z)$ \footnote{$S_8^\times(z)$ constraints are not to be confused with the redshift-independent constraints denoted in this paper as $S_8^\times$ which are defined at $z=0$.} for each of our redshift bin means, defined as the following:
\begin{equation}
    S_8^\times(z) = \sigma_8(z) \left(\dfrac{\Omega_m(z = 0)}{0.3}\right)^{0.4}
\end{equation}
which is a rescaling of $S_8^\times(z=0)$ measured from each redshift bin, computed by assuming the \Planck~PR3 fiducial cosmology \cite{Planck_2018_VI} to evaluate the matter power spectrum and the linear growth factor $D(z)$ using \texttt{CAMB} (\cite{Lewis_1999bs}, \cite{Lewis_2002ah}, \cite{Howlett_2012mh}) for both $\Omega_m(z=0)$ and $\sigma_8(z)$:
\begin{equation}
    \sigma_8(z) = D(z) \, \sigma_8^{\rm PR3}(z=0).
\end{equation} This leads to the implication that if our parameter of structure formation at the present-day is in agreement with \Planck, our structure growth amplitude should scale using this function of redshift with the same behavior shown by \Planck. These rescaled, redshift-dependent constraints are shown in Figure \ref{fig:s8xz}.

\subsection{Parameter recovery tests}
\label{sec:param-recovery}
To ensure we are robust to biases from ``prior volume" effects, where the posterior mean is found to deviate from the maximum a posteriori value due to the influence of a number of prior-dominated parameters, we perform parameter recovery tests in which we attempt to recover exactly known input cosmological parameters using noiseless theory spectra computed using those same exact parameters. 

To do this, we bin a set of noiseless theory spectra in the same way as our measurement's data bandpowers are binned (see Section \ref{sec:angular-power-spectrum}), and pass that into our MCMC sampler as the data vector. We use our joint hybrid covariance matrix described in Section \ref{sec:covmat} that contains information from the ACT DR6 x DESI LRG cross-correlation, the \Planck~PR4 x DESI LRG cross-correlation, and the DESI LRG auto-correlation spectra. This parameter recovery test also allows us to measure the $\Omega_m$ dependence on this paper's headline result, which is the combination of $\sigma_8$ and $\Omega_m$ with the lowest relative error -- we compute this to be:
\begin{equation}
    S_8^\times = \sigma_8 \left(\dfrac{\Omega_m}{0.3}\right)^{0.4}
\end{equation}

Using the Buzzard simulations and their associated cosmological parameters as inputs to generate noiseless theory spectra, the parameter recovery test allows us to recover $S_8^\times$ to within less than 0.1 $\sigma$ from the input value (considering both the posterior mean as well as the best-fit) for the baseline joint \Planck~+ ACT analysis when combining all redshift bins and under 0.4 $\sigma$ for different combinations of \Planck~and ACT with individual redshift bins. 

This test is not to be confused with a similar systematics test of fitting to the Buzzard simulations' data vector, which is noisy and computed using a simulated CMB lensing convergence field intrinsic to the simulation suite. This test confirms the robustness of the theory model and also acts as a robustness check for the appropriate bandpower window functions, pixel window function, analysis covariance matrices, and choice of priors. Further details of this test and adequate recoveries of $S_8$ and $\sigma_8$ with and without a BAO prior are described in Section 5.5 in \cite{Sailer_2024}, where $S_8^\times$ is recovered and constrained to a posterior mean and best-fit value less than 0.5 $\sigma$ from truth for all combinations of redshift bins and covariance matrices.

\subsection{Results}
\label{sec:results}

\begin{figure*}
\centering
\includegraphics[width=\textwidth]{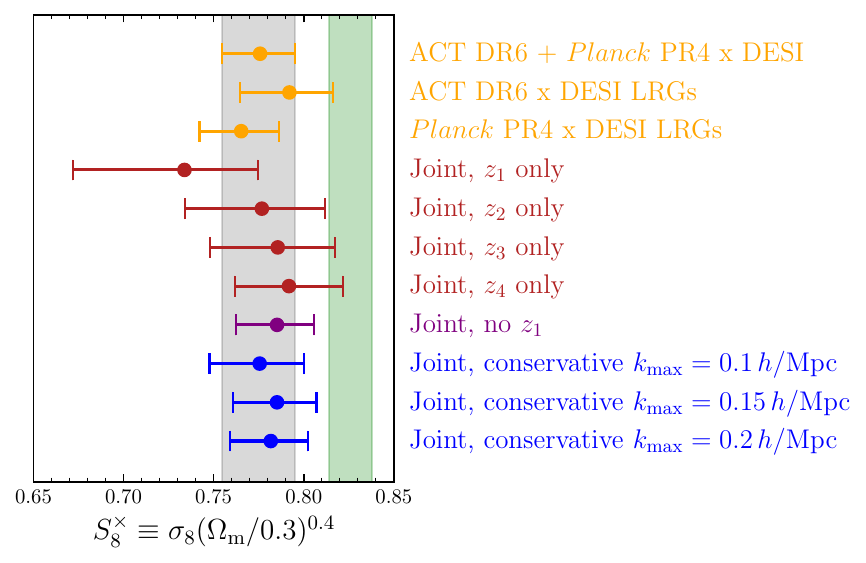}
\caption{We show $S_8^\times$ constraints using different analysis variations and demonstrate their consistency with the baseline constraints, shown in yellow for ACT + \Planck, ACT only, and \Planck~only respectively. The red points show the joint ACT DR6 and \Planck~PR4 constraints when fit to each redshift bin independently; seeing that the mean of the lowest redshift bin lies outside the 1$\sigma$ range of our baseline constraint (a feature of this redshift bin we also see with the ACT-only and Planck-only cross-correlations in Figure \ref{fig:s8xz}), we verify that the combination of the other 3 redshift bins (shown in purple) is consistent with our baseline mean. The blue points show our analysis carried out with the conservative scale cuts shown in Table \ref{tab:kmax-snr-percent}. For reference, we also show a green band representing the constraint from the \Planck~PR4 primary CMB anisotropies \cite{Rosenberg_2022}.}
\label{fig:variations}
\end{figure*}

Combining the posterior information from the ACT DR6 x DESI LRG cross-correlation
power spectrum and DESI LRG auto-correlation power spectrum, we have (with best-fit values in brackets):
\begin{equation}
S_8^\times \text{[\textit{DR6}]} = 0.792^{+0.024}_{-0.028} \,\,[0.797]
\end{equation}
The combination of the \Planck~PR4 x DESI LRG cross-correlation power spectrum and the DESI LRG auto-correlation power spectrum gives us a slightly tighter constraint albeit a lower mean:
\begin{equation}
S_8^\times \text{[\textit{PR4}]} = 0.766 \pm 0.022\,\, [0.769]
\end{equation}
Our baseline results use the combination of ACT and \Planck~cross-correlations with DESI, which yields this analysis's strongest constraint at 2.7\%:
\begin{equation}
S_8^\times \equiv \sigma_8 \left( \dfrac{\Omega_m}{0.3}\right)^{0.4} =0.776^{+0.019}_{-0.021} \,\, [0.776]
\end{equation}
a result that is approximately 2.1$\sigma$\footnote{Here and throughout the paper, we define a difference or discrepancy between measurement $\mu_1 \pm \sigma_1$ and measurement $\mu_2 \pm \sigma_2$ as $(\mu_1 - \mu_2) / \sqrt{\sigma_1^2 + \sigma_2^2}$. For a posterior constraint with asymmetric error bars $\mu^{+x}_{-y}$, we compute and quote differences using the standard deviation of the samples used to construct the posterior.} lower (1.2$\sigma$ lower for ACT only) than the \Planck~PR4 prediction of:
\begin{equation}
    S_8^\times = 0.826 \pm 0.012
\end{equation} from the primary CMB anisotropies (2.2$\sigma$ lower than \Planck~PR3), while being in general agreement with the late-time galaxy lensing constraints. In all three cases we see no significant tension between the best-fit values and the posterior means, showing that we are not affected by prior volume effects on $S_8^\times$.  A feature of these results is that, as seen in Figure \ref{fig:variations}, the constraint from using only the lowest redshift bin is more than 0.5$\sigma$ low from the baseline constraint mean, an effect that was similarly observed in \cite{White_2022} but to a greater extent than our analysis's findings; this effect is not specific to the \Planck-only cross-correlation as the ACT-only constraint for this redshift bin presents a similar discrepancy from the \Planck~ primary CMB. This lowest redshift bin is the least constraining and features the largest error bars of the four redshift bins. We proceeded to run a joint constraint while excluding this lowest redshift bin, which pushes our $S_8^\times$ mean to a slightly higher value ($S_8^\times = 0.785^{+0.021}_{-0.023}$) but not high enough to be in tension ($< 0.3\sigma$) with our baseline result. We also run a set of varied constraints where the maximum multipole scale cuts are more conservative and reflect the maximum $k$ and $L$ scale cuts shown in Figure \ref{tab:kmax-snr-percent}; these results are still consistent with our baseline constraint on $S_8^\times$, showing that our analysis is robust to different scale cuts. Further details on a thorough test of our consistency with a linear theory model and a ``model independent" approach can be seen in our companion paper \cite{Sailer_2024}. 

\begin{table}
\centering
\begin{tabular} { l  c  c  c  c  c}
    $k_{\rm max}$ & $L_{\rm max} \, (z_1 \rightarrow z_4)$ & DR6 & PR4 & DR6 + PR4 & $S_8^{\times}$ \% constraint \\
   \hline \hline
   $0.1 \,\, h / \text{Mpc}$ & 124, 124, 178, 178 & 22 & 23 & 31 & 3.3\%\\
   $0.15 \,\, h / \text{Mpc}$ & 178, 178, 243, 317 & 28 & 29 & 38 & 2.9\%\\
   $0.2 \,\, h / \text{Mpc}$ & 243, 317, 317, 401 & 31 & 33 & 43 & 2.7\%\\
   \textbf{Baseline} ($0.5 \, h / \text{Mpc}$)& 600, 600, 600, 600 & \textbf{38} & \textbf{39} & \textbf{50} & \textbf{2.7\%} \\
   \hline
\end{tabular}
\caption{This table shows the signal-to-noise ratio (computed as $\sqrt{\chi^2}$, see Appendix \ref{sec:snr-calculation}) of the $C_L^{\kappa g}$ measurement with ACT DR6 lensing, \Planck~PR4 lensing, and the joint ACT + \Planck~analysis; the corresponding strongest percentage constraint of $S_8^\times$ inferred from their respective posteriors are shown in the right-most column, each shown with its dependence on the maximum scale wavenumber, $k_{\rm max}$. For each redshift bin, we relate this $k_{\rm max}$ to the maximum angular multipole $L_{\rm max}$ using the comoving distance corresponding to the peak of the redshift distribution, and use $L_{\rm max}$ to determine the scales in the covariance matrix and fiducial theory bandpowers used to compute $\chi^2$. The results show us also how much improvement we gain in our fractional constraint and SNR by using HEFT and smaller scales compared to a linear theory-like model (first three entries).}
\label{tab:kmax-snr-percent}
\end{table}

We show $S_8^\times(z)$ for each of our 4 redshift bins in Figure \ref{fig:s8xz} by rescaling the ACT + \Planck~$S_8^\times$ means and errors from redshift zero to their effective redshifts (see Table 1 in \cite{Sailer_2024}); curly bracketed values represent the \Planck~PR4 \cite{Rosenberg_2022} primary CMB prediction of $S_8^\times(z)$:
\begin{align*}
    S_8^\times(z = 0.470) &= 0.572^{+0.032}_{-0.048} \,\,\{0.641\}\\
    S_8^\times(z = 0.625) &= 0.560^{+0.025}_{-0.031} \,\,\{0.594\}\\
    S_8^\times(z = 0.785) &= 0.525^{+0.021}_{-0.025} \,\,\{0.550\}\\
    S_8^\times(z = 0.914) &= 0.498 \pm 0.019 \,\,\{0.518\}
\end{align*}
and note that the first redshift bin as we found previously shows the lowest mean compared to the primary CMB prediction. 
As demonstrated in the companion paper \cite{Sailer_2024}, these $S_8^\times(z)$ values are correlated with each other by approximately 0–30\%, with higher correlations found between redshift bins 3 and 4 (that we also observe in Figure \ref{fig:joint_covmat}); an optimal linear combination of the $S_8^\times(z)$ constraints weighted by their respective correlations recovers the baseline joint constraint to $<0.1 \sigma$, confirming our lower value with \Planck~PR3 at the 2.2 $\sigma$ significance level.

\begin{figure*}
\centering
\includegraphics[width=\columnwidth]{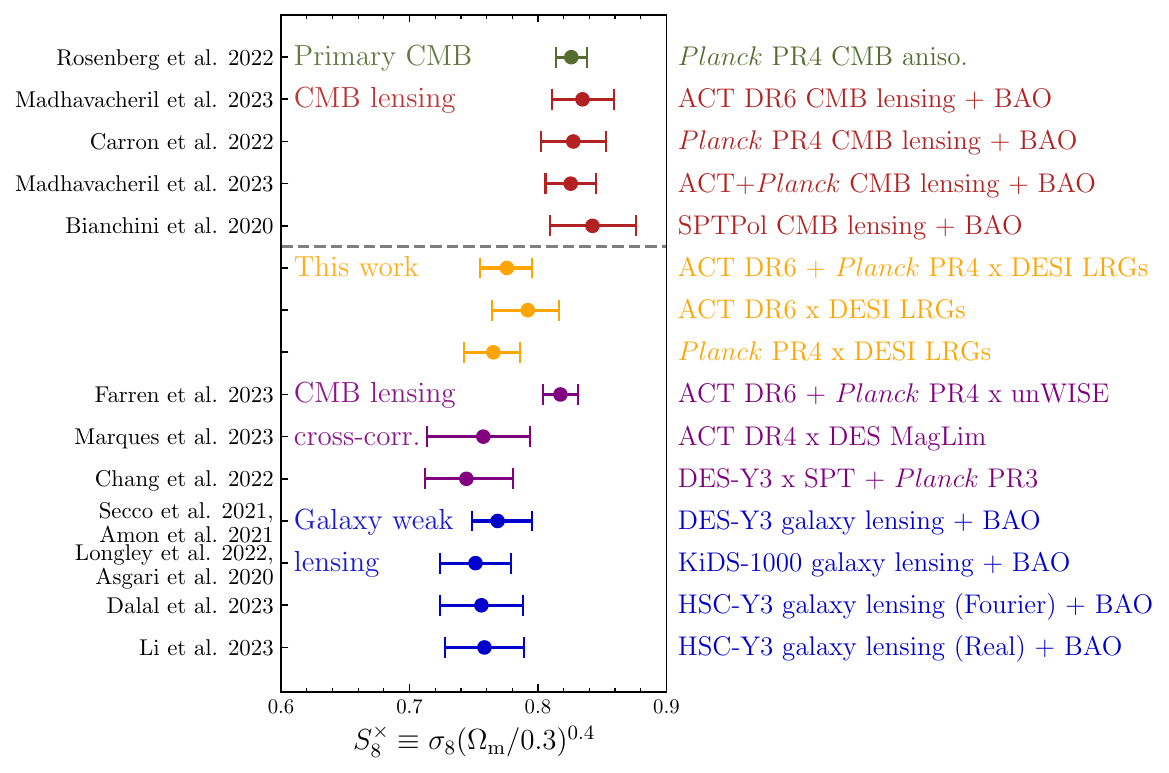}
\caption{We show our constraint on $S_8^\times$ with the ACT DR6 cross-correlation, \Planck~PR4 cross-correlation, and the joint ACT and \Planck~analysis in yellow. We find that we are not in substantial disagreement with constraints from the primary CMB (in green, \cite{Rosenberg_2022}), CMB lensing power spectrum (in red, \cite{madhavacheril2023atacama, Carron_2022, Bianchini_2020}), and from galaxy weak lensing (in blue, \cite{Secco_2021, Amon_2021, Longley_2023, Amon_2022, Dalal_2023, Li_2023}). We have various levels of agreement with different galaxy-CMB lensing cross-correlations (in purple, \cite{farren2023atacama, Marques_2024, White_2022, Chang_2023}) and lower redshift tracers (such as DES \texttt{MagLim} \cite{Marques_2024}, unWISE Green, Blue  \cite{farren2023atacama}).}
\label{fig:s8x_comparison}
\end{figure*}

\newpage

\section{Summary and Discussion}

Through a harmonic-space tomographic cross-correlation between state-of-the-art CMB lensing maps from \Planck~and ACT with DESI LRGs, we have obtained a 2.7\% constraint on the parameter combination $S_8^{\times}\equiv \sigma_8 (\Omega_m/0.3)^{0.4}$ characterizing the amplitude of matter fluctuations. As seen in Figure \ref{fig:s8x_comparison}, our ACT-only constraint of $S_8^\times = 0.792^{+0.024}_{-0.028}$ and our joint ACT + \Planck~constraint of $S_8^{\times} = 0.776^{+0.019}_{-0.021}$ are roughly consistent both with the \Planck~PR4 primary CMB anisotropy prediction (our ACT-only and joint constraints are lower by 1.2$\sigma$ and 2.1$\sigma$ respectively) as well as with other large scale structure (LSS) constraints (which generally come lower than the \Planck~prediction by $1-2.5\sigma$). An open question is whether the mild discrepancy of several of these LSS probes is driven by new physics, unaccounted astrophysical processes (e.g., baryonic feedback), systematics, or statistical fluctuations. Since every probe has sensitivity to different scales and redshifts, high-precision cross-correlations such as from this work bring us closer to clarifying the origin of these discrepancies. 

This cross-correlation result computed for four redshift bins is robust and verified to be not significantly biased by extragalactic or Galactic foregrounds as well as other systematics. We demonstrate this using the LRG-like HOD cross-correlation test with the Websky foregrounds-only reconstruction along with our comprehensive suite of 48 null tests using a variety of ACT DR6 lensing products in which we see no significant spurious correlations with expected null signals. We follow a blinding procedure to avoid the influence of confirmation bias, and ensure that the analysis design choices including the HEFT theory modeling, multipole scale cuts, ``hybrid'' covariance matrix, and likelihood / prior parameterization are devised and fixed before comparing and fitting our theory model to the unblinded data.

Generally, constraints from the CMB lensing auto-spectrum (which probe predominantly higher redshifts $z=1-2$ and linear scales $k<0.2 \,{\rm Mpc}^{-1}$) show excellent agreement with the \Planck~CMB prediction. At the same time, cross-correlations of CMB lensing with unWISE (probing $z\sim 0.6$ and $z\sim 1.1$) are also consistent with \Planck.  We have explored the redshift dependence of our $S_8^{\times}$ constraint by separately constraining this parameter for each redshift bin. While all bins remain nominally consistent with \Planck, the lowest redshift bin shows the largest difference in the mean $S_8^\times$ value; an analysis that excludes this redshift bin is consistent with \Planck~at 1.6$\sigma$. This may be an indication of new physics (e.g., modified gravity), a systematic that affects lower redshifts more, or a statistical fluctuation, though no strong conclusion can be drawn given the uncertainty on our lowest redshift bin.  Future CMB lensing cross-correlations with the DESI Legacy Imaging galaxies \cite{Hang_2020, Qu_2024},  DESI Bright Galaxy Sample (BGS) and other lower redshift samples will be key to assessing this conclusively. These cross-correlations will also be significantly improved in precision with future CMB lensing surveys such as the Simons Observatory (SO, \cite{Ade_2019}), CMB-Stage 4 (CMB-S4, \cite{cmbs4sciencebook}), and CMB-HD (\cite{macinnis2024cmbhd}), allowing for a path to disentangling possible discrepancies between early-time and late-time observations of structure formation.

\acknowledgments
\input{acknowledgements.tex}

\appendix
\section{Null test plots}
\label{sec:appendix-null-tests}
\begin{figure*}[!htb]
\centering
\includegraphics[width=0.495\columnwidth]{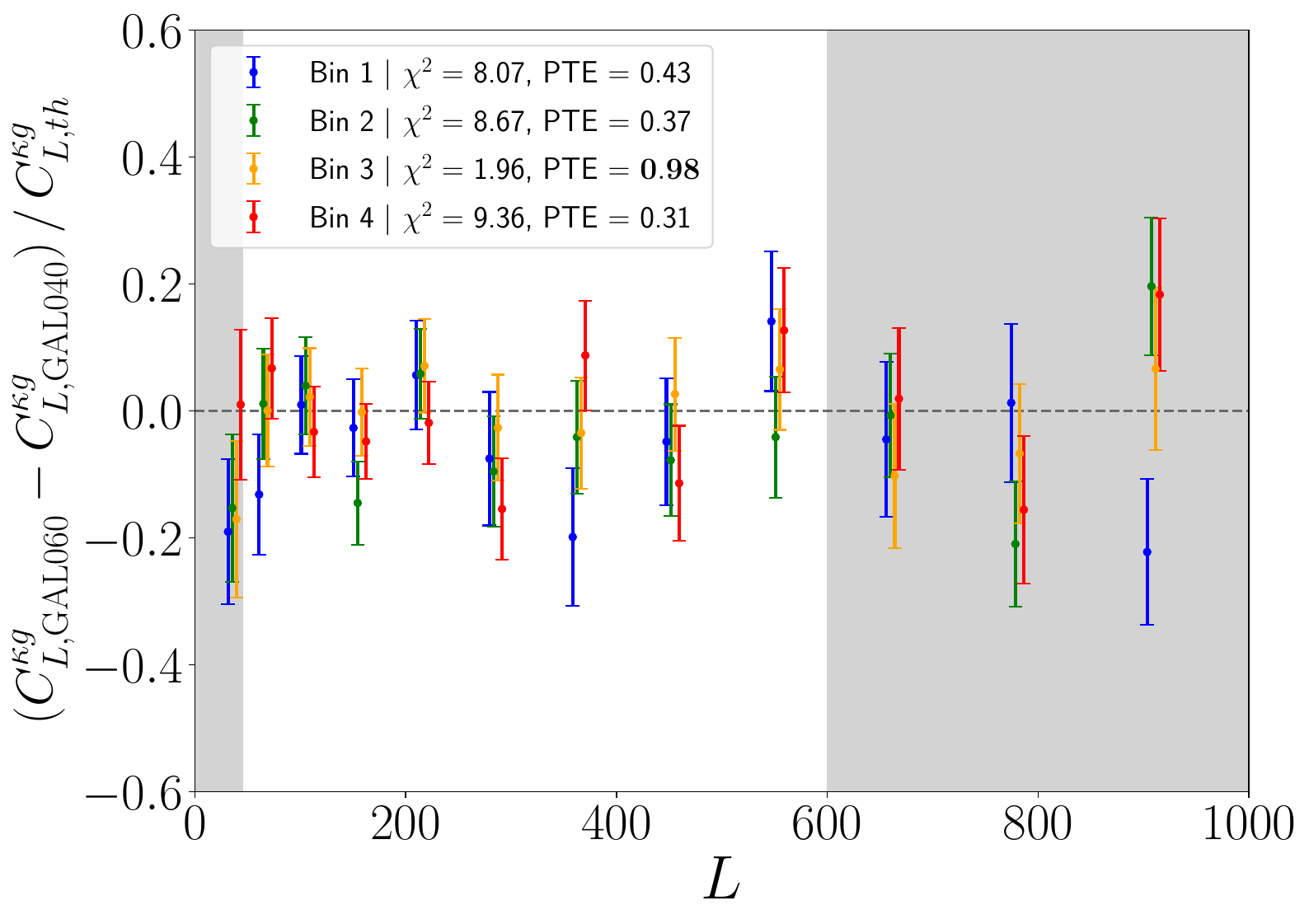}
\includegraphics[width=0.495\columnwidth]{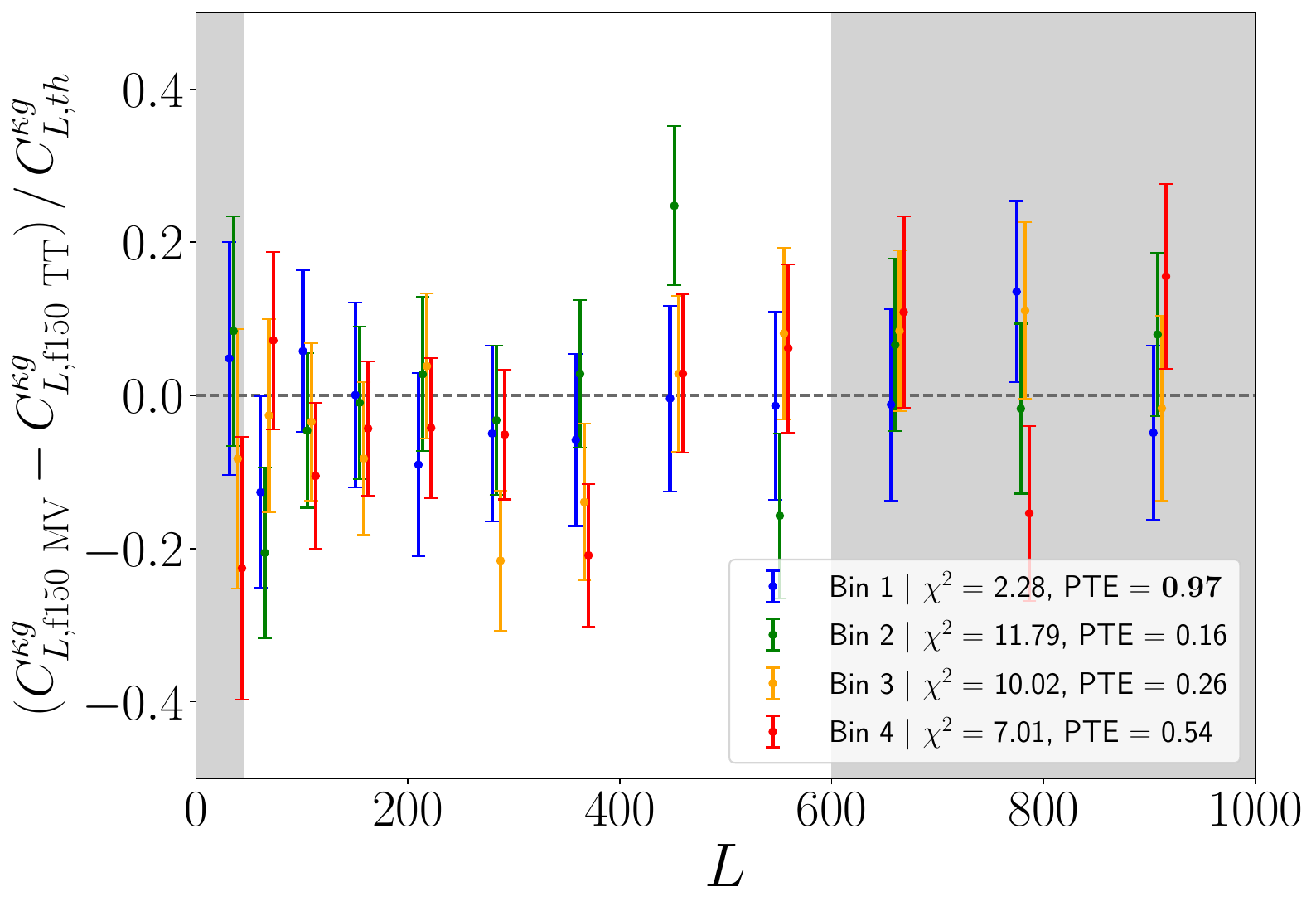}
\caption{We show two examples of bandpower-level null test ``failures'' here. \textit{Left: } The ACT DR6 lensing convergence map is cross-correlated with a galaxy redshift bin using two versions of the lensing analysis mask, \texttt{GAL060} (mask used for baseline analysis) and \texttt{GAL040} (more restrictive mask used for extended Galactic foreground mitigation). The difference of their respective spectra is shown here, with bin 3 failing due to a high PTE. \textit{Right: } Here, we only consider the \texttt{f150} CMB split, and reconstruct lensing from it separately using the $MV$ and $TT$-only quadratic estimators, and take the difference of their respective spectra. Bin 1 ultimately fails this test due to a high PTE. More details on all of these tests can be seen in Section \ref{sec:nulls}.}
\label{fig:null_test_extra_1}
\end{figure*}

\begin{figure*}[!htb]
\includegraphics[width=0.495\columnwidth]{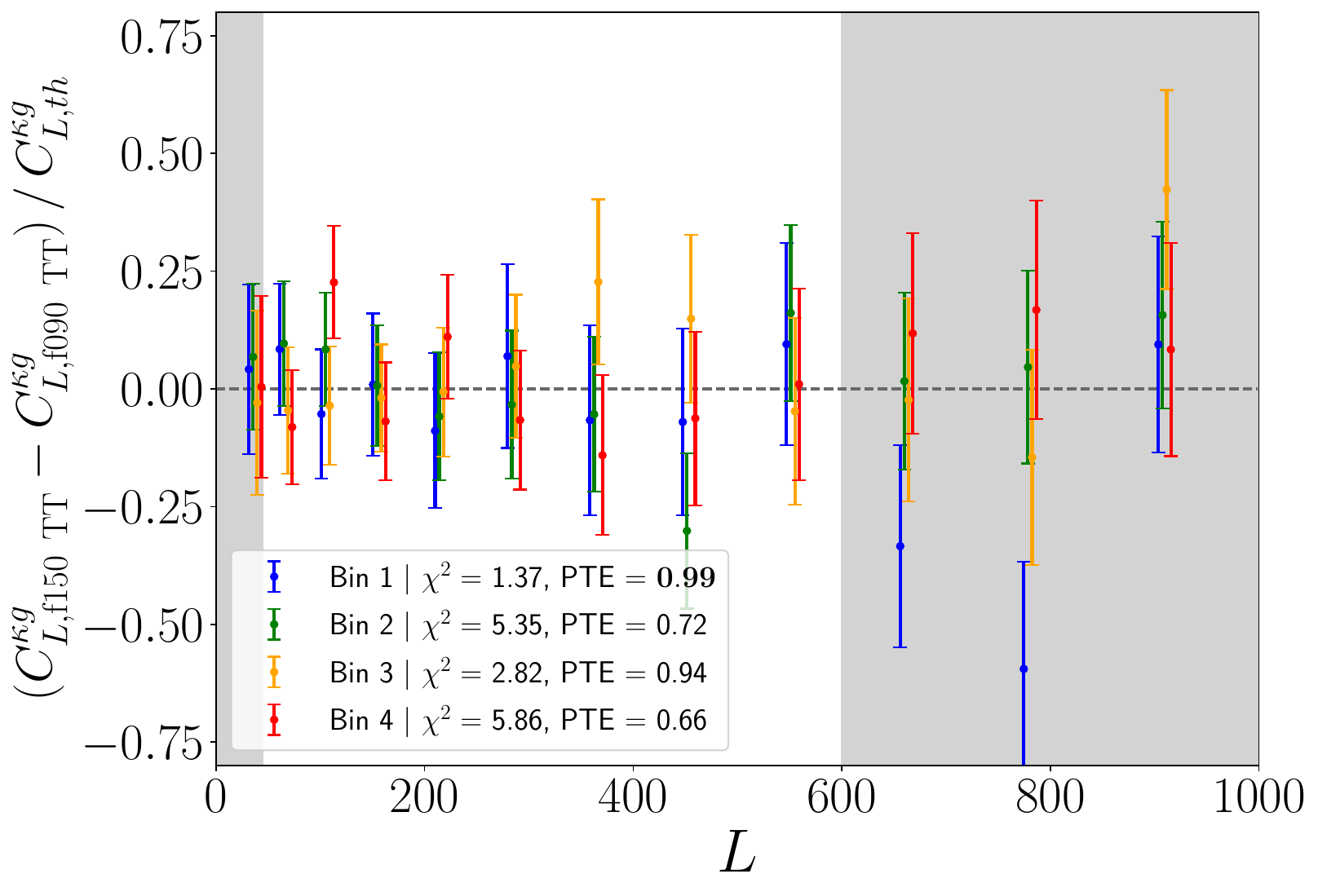}
\includegraphics[width=0.495\columnwidth]{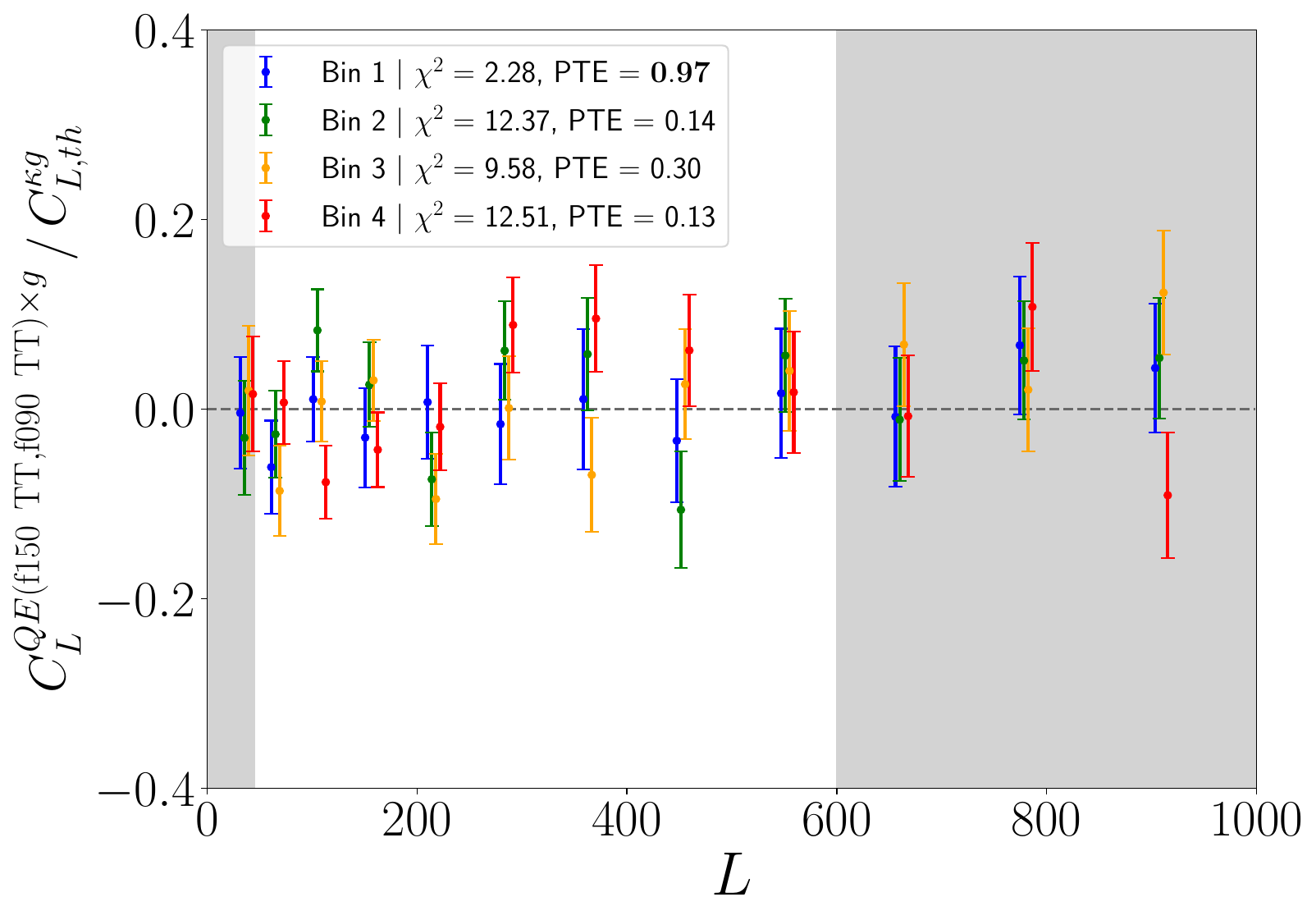}
\caption{We show the results of a null test computed with the \texttt{f150} CMB split and the \texttt{f090} CMB split, with different combinations passed into a $TT$-only CMB lensing reconstruction. The fact that both of these tests ``fail'' for using the same data products for the same redshift bin (Bin 1) shows that these outcomes are likely to be correlated. \textit{Left: } The difference is computed after each CMB split is passed into the reconstruction pipeline, with each reconstructed convergence cross-correlated with a galaxy redshift bin and subtracted as spectra (bandpower-level). \textit{Right: } The difference is computed before passing the data product into the reconstruction pipeline, with the lensing signal reconstructed from the map difference of the CMB splits and then cross-correlated with a galaxy redshift bin (map-level). More details on all of these tests can be seen in Section \ref{sec:nulls}.}
\label{fig:null_test_extra_2}
\end{figure*}
Here in Figures \ref{fig:null_test_extra_1} and \ref{fig:null_test_extra_2} we display plots of the remaining null tests not shown in the main body of the paper that result in failures defined by the criteria set in Section \ref{sec:nulls}.

\section{SNR calculation}
\label{sec:snr-calculation}
The signal-to-noise ratio for a measurement of $C_L^{\kappa g}$ can be simply expressed as:
\begin{equation}
\text{SNR} = \sqrt{\displaystyle\sum_{L} \text{SNR}^2(L)} = \sqrt{\displaystyle\sum_{L} \dfrac{(C_L^{\kappa g})^2}{\sigma^2\left(C_L^{\kappa g}\right)}} \\
\label{eqn:snr-basic}
\end{equation}
For our purposes, we may wish to compare SNR values across different sets of data, spectra, and covariances, so $C_L^{\kappa g}$ in the numerator of equation \ref{eqn:snr-basic} is usually represented by an invariant fiducial theory spectrum while the $\sigma$'s may change depending on the error bars placed on a specific measurement.

However, summing over each bandpower independently ignores correlations between bandpowers, so one takes into account $\mathcal{C}$, the covariance matrix block for $\textbf{d} = C_L^{\kappa g}$, in lieu of the latter expression:
\begin{equation}
    \text{SNR} = \sqrt{\textbf{d}^T \cdot \mathcal{C}^{-1} \cdot \textbf{d}} \equiv \sqrt{\chi^2 (C_L^{\kappa g})}
\end{equation}
where the cumulative SNR for a multipole range of $[L_{\rm min}, L_{\rm max}]$ can be expressed as:
\begin{equation}
    \text{SNR}(L_{\rm min}, L_{\rm max}) = \left[ \displaystyle\sum_{L = L_{\rm min}}^{L_{\rm max}} \displaystyle\sum_{L' = L_{\rm min}}^{L_{\rm max}} C_L^{\kappa g} \times \text{Cov}\left(C_L^{\kappa g}, C_{L'}^{\kappa g}\right)^{-1} \times C_{L'}^{\kappa g} \right]^{1/2}
    \label{eqn:snr-cumulative}
\end{equation}
To compute the contribution to the SNR per bandpower, we compute the following for a given binning scheme of bin edges $L_i \in [L_0\, (= L_{\rm min}), L_1, \ldots, L_{\rm max}]$:
\begin{equation}
    \text{SNR}(L_i, L_{i+1}) = \sqrt{\text{SNR}^2(L_{\rm min}, L_{i+1}) - \text{SNR}^2(L_{\rm min}, L_i)}
\end{equation}
where the right side of this equation is computed using Equation \ref{eqn:snr-cumulative}. This value is plotted for each analysis multipole bin in Figure \ref{fig:clkk} after applying an arbitrary normalization factor, and the relative fraction of the SNR that each bandpower contributes can be calculated by dividing the value for each bin by the baseline total SNR found in Table \ref{tab:kmax-snr-percent}. 

\newpage
\bibliographystyle{JHEP}
\bibliography{biblio}
\appendix

\end{document}

%% file: authors_kim.tex
\author[1]{Joshua Kim\,\orcidlink{0000-0002-0935-3270}}
\author[2,3]{Noah Sailer\,\orcidlink{0000-0002-5333-8983}}
\author[1]{Mathew S. Madhavacheril}
\author[3,2]{Simone Ferraro\,\orcidlink{0000-0003-4992-7854}}
\author[4,5]{Irene~Abril-Cabezas\,\orcidlink{0000-0003-3230-4589}}
\author[6]{Jessica Nicole Aguilar}
\author[7]{Steven~Ahlen\,\orcidlink{0000-0001-6098-7247}}
\author[8]{J.~Richard~Bond}
\author[9]{David~Brooks}
\author[10]{Etienne~Burtin}
\author[11]{Erminia Calabrese}
\author[12]{Shi-Fan~Chen}
\author[13]{Steve~K.~Choi\,\orcidlink{0000-0002-9113-7058}}
\author[6]{Todd~Claybaugh}
\author[14]{Omar Darwish}
\author[15]{Axel~de la Macorra\,\orcidlink{0000-0002-1769-1640}}
\author[6]{Joseph~DeRose}
\author[1]{Mark Devlin\,\orcidlink{0000-0002-3169-9761}}
\author[16]{Arjun~Dey\,\orcidlink{0000-0002-4928-4003}}
\author[9]{Peter~Doel}
\author[17,18]{Jo~Dunkley\,\orcidlink{0000-0002-7450-2586}}
\author[4]{Carmen~Embil-Villagra\,\orcidlink{0009-0001-3987-7104}}
\author[4,5]{Gerrit~S.~Farren}
\author[9,19]{Andreu~Font-Ribera\,\orcidlink{0000-0002-3033-7312}}
\author[20,21]{Jaime~E.~Forero-Romero\,\orcidlink{0000-0002-2890-3725}}
\author[22,23,24]{Enrique~Gaztañaga}
\author[25]{Vera Gluscevic\,\orcidlink{0000-0002-3589-8637}}
\author[6]{Satya~Gontcho A Gontcho\,\orcidlink{0000-0003-3142-233X}}
\author[6]{Julien~Guy\,\orcidlink{0000-0001-9822-6793}}
\author[26,27,28]{Klaus~Honscheid}
\author[29]{Cullan~Howlett\,\orcidlink{0000-0002-1081-9410}}
\author[30]{David~Kirkby\,\orcidlink{0000-0002-8828-5463}}
\author[6]{Theodore~Kisner\,\orcidlink{0000-0003-3510-7134}}
\author[6]{Anthony~Kremin\,\orcidlink{0000-0001-6356-7424}}
\author[6]{Martin~Landriau\,\orcidlink{0000-0003-1838-8528}}
\author[31]{Laurent~Le~Guillou\,\orcidlink{0000-0001-7178-8868}}
\author[6]{Michael~E.~Levi\,\orcidlink{0000-0003-1887-1018}}
\author[4,5]{Niall MacCrann\,\orcidlink{0000-0002-8998-3909}}
\author[32,19]{Marc~Manera\,\orcidlink{0000-0003-4962-8934}}
\author[33,34]{Gabriela~A.~Marques\,\orcidlink{0000-0002-8571-8876}}
\author[16]{Aaron~Meisner\,\orcidlink{0000-0002-1125-7384}}
\author[35,19]{Ramon~Miquel}
\author[36,37]{Kavilan Moodley\,\orcidlink{0000-0001-6606-7142}}
\author[38]{John~Moustakas\,\orcidlink{0000-0002-2733-4559}}
\author[39]{Laura~B.~Newburgh}
\author[40]{Jeffrey~A.~Newman\,\orcidlink{0000-0001-8684-2222}}
\author[41,42]{Gustavo~Niz\,\orcidlink{0000-0002-1544-8946}}
\author[1]{John Orlowski-Scherer\,\orcidlink{0000-0003-1842-8104}}
\author[10,6]{Nathalie~Palanque-Delabrouille\,\orcidlink{0000-0003-3188-784X}}
\author[43,44,45]{Will~J.~Percival\,\orcidlink{0000-0002-0644-5727}}
\author[46]{Francisco~Prada\,\orcidlink{0000-0001-7145-8674}}
\author[4,5,47]{Frank~J.~Qu\,\orcidlink{0000-0001-7805-1068}}
\author[48]{Graziano~Rossi}
\author[49]{Eusebio~Sanchez\,\orcidlink{0000-0002-9646-8198}}
\author[50,51]{Emmanuel Schaan}
\author[52]{Edward~F.~Schlafly\,\orcidlink{0000-0002-3569-7421}}
\author[6]{David~Schlegel}
\author[53,54]{Michael~Schubnell}
\author[55]{Neelima Sehgal\,\orcidlink{0000-0002-9674-4527 }}
\author[56]{Hee-Jung~Seo\,\orcidlink{0000-0002-6588-3508}}
\author[57]{Shabbir Shaikh\,\orcidlink{0000-0001-6731-0351}}
\author[4,5]{Blake~D.~Sherwin}
\author[58]{Crist\'obal Sif\'on\,\orcidlink{0000-0002-8149-1352}}
\author[16]{David~Sprayberry}
\author[17]{Suzanne~T.~Staggs\,\orcidlink{0000-0002-7020-7301}}
\author[54]{Gregory~Tarl\'{e\,}\,\orcidlink{0000-0003-1704-0781}}
\author[57]{Alexander van Engelen\,\orcidlink{0000-0002-3495-158X}}
\author[16]{Benjamin~Alan~Weaver}
\author[59]{Lukas Wenzl}
\author[2,3]{Martin White}
\author[60]{Edward J. Wollack\,\orcidlink{0000-0002-7567-4451}}
\author[10]{Christophe~Yèche\,\orcidlink{0000-0001-5146-8533}}
\author[61]{Hu~Zou\,\orcidlink{0000-0002-6684-3997}}

\affiliation[1]{Department of Physics and Astronomy, University of Pennsylvania, 209 South 33rd Street, Philadelphia, PA 19104, USA}
\affiliation[2]{Department of Physics, University of California, Berkeley, CA 94720, USA}
\affiliation[3]{Lawrence Berkeley National Laboratory, One Cyclotron Road, Berkeley, CA 94720, USA}
\affiliation[4]{DAMTP, Centre for Mathematical Sciences, University of Cambridge, Wilberforce Road, Cambridge CB3 OWA, UK}
\affiliation[5]{Kavli Institute for Cosmology Cambridge, Madingley Road, Cambridge CB3 0HA, UK}
\affiliation[6]{Lawrence Berkeley National Laboratory, 1 Cyclotron Road, Berkeley, CA 94720, USA}
\affiliation[7]{Physics Dept., Boston University, 590 Commonwealth Avenue, Boston, MA 02215, USA}
\affiliation[8]{Canadian Institute for Theoretical Astrophysics, 60 St. George Street, University of Toronto, Toronto, ON, M5S 3H8, Canada}
\affiliation[9]{Department of Physics \& Astronomy, University College London, Gower Street, London, WC1E 6BT, UK}
\affiliation[10]{IRFU, CEA, Universit\'{e} Paris-Saclay, F-91191 Gif-sur-Yvette, France}
\affiliation[11]{School of Physics and Astronomy, Cardiff University, The Parade, Cardiff, Wales CF24 3AA, UK}
\affiliation[12]{School of Natural Sciences, Institute for Advanced Study, 1 Einstein Drive, Princeton, NJ 08540, USA}
\affiliation[13]{Department of Physics and Astronomy, University of California, Riverside, CA 92521, USA}
\affiliation[14]{Université de Genève, Département de Physique Théorique et CAP, 24 Quai Ansermet, CH-1211 Genève 4, Switzerland}
\affiliation[15]{Instituto de F\'{\i}sica, Universidad Nacional Aut\'{o}noma de M\'{e}xico,  Cd. de M\'{e}xico  C.P. 04510,  M\'{e}xico}
\affiliation[16]{NSF NOIRLab, 950 N. Cherry Ave., Tucson, AZ 85719, USA}
\affiliation[17]{Joseph Henry Laboratories of Physics, Jadwin Hall, Princeton University, Princeton, NJ, USA 08544}
\affiliation[18]{Department of Astrophysical Sciences, Peyton Hall, Princeton University, Princeton, NJ USA 08544}
\affiliation[19]{Institut de F\'{i}sica d’Altes Energies (IFAE), The Barcelona Institute of Science and Technology, Campus UAB, 08193 Bellaterra Barcelona, Spain}
\affiliation[20]{Departamento de F\'isica, Universidad de los Andes, Cra. 1 No. 18A-10, Edificio Ip, CP 111711, Bogot\'a, Colombia}
\affiliation[21]{Observatorio Astron\'omico, Universidad de los Andes, Cra. 1 No. 18A-10, Edificio H, CP 111711 Bogot\'a, Colombia}
\affiliation[22]{Institut d'Estudis Espacials de Catalunya (IEEC), 08034 Barcelona, Spain}
\affiliation[23]{Institute of Cosmology and Gravitation, University of Portsmouth, Dennis Sciama Building, Portsmouth, PO1 3FX, UK}
\affiliation[24]{Institute of Space Sciences, ICE-CSIC, Campus UAB, Carrer de Can Magrans s/n, 08913 Bellaterra, Barcelona, Spain}
\affiliation[25]{Department of Physics and Astronomy, University of Southern California, Los Angeles, CA 90089, USA}
\affiliation[26]{Center for Cosmology and AstroParticle Physics, The Ohio State University, 191 West Woodruff Avenue, Columbus, OH 43210, USA}
\affiliation[27]{Department of Physics, The Ohio State University, 191 West Woodruff Avenue, Columbus, OH 43210, USA}
\affiliation[28]{The Ohio State University, Columbus, 43210 OH, USA}
\affiliation[29]{School of Mathematics and Physics, University of Queensland, 4072, Australia}
\affiliation[30]{Department of Physics and Astronomy, University of California, Irvine, 92697, USA}
\affiliation[31]{Sorbonne Universit\'{e}, CNRS/IN2P3, Laboratoire de Physique Nucl\'{e}aire et de Hautes Energies (LPNHE), FR-75005 Paris, France}
\affiliation[32]{Departament de F\'{i}sica, Serra H\'{u}nter, Universitat Aut\`{o}noma de Barcelona, 08193 Bellaterra (Barcelona), Spain}
\affiliation[33]{Fermi National Accelerator Laboratory, P. O. Box 500, Batavia, IL 60510, USA}
\affiliation[34]{Kavli Institute for Cosmological Physics, University of Chicago, Chicago, IL 60637, USA}
\affiliation[35]{Instituci\'{o} Catalana de Recerca i Estudis Avan\c{c}ats, Passeig de Llu\'{\i}s Companys, 23, 08010 Barcelona, Spain}
\affiliation[36]{Astrophysics Research Centre, University of KwaZulu-Natal, Westville Campus, Durban 4041, South Africa}
\affiliation[37]{School of Mathematics, Statistics \& Computer Science, University of KwaZulu-Natal, Westville Campus, Durban 4041, South Africa}
\affiliation[38]{Department of Physics and Astronomy, Siena College, 515 Loudon Road, Loudonville, NY 12211, USA}
\affiliation[39]{Yale University, Department of Physics, New Haven, CT, 06511}
\affiliation[40]{Department of Physics \& Astronomy and Pittsburgh Particle Physics, Astrophysics, and Cosmology Center (PITT PACC), University of Pittsburgh, 3941 O'Hara Street, Pittsburgh, PA 15260, USA}
\affiliation[41]{Departamento de F\'{i}sica, Universidad de Guanajuato - DCI, C.P. 37150, Leon, Guanajuato, M\'{e}xico}
\affiliation[42]{Instituto Avanzado de Cosmolog\'{\i}a A.~C., San Marcos 11 - Atenas 202. Magdalena Contreras, 10720. Ciudad de M\'{e}xico, M\'{e}xico}
\affiliation[43]{Department of Physics and Astronomy, University of Waterloo, 200 University Ave W, Waterloo, ON N2L 3G1, Canada}
\affiliation[44]{Perimeter Institute for Theoretical Physics, 31 Caroline St. North, Waterloo, ON N2L 2Y5, Canada}
\affiliation[45]{Waterloo Centre for Astrophysics, University of Waterloo, 200 University Ave W, Waterloo, ON N2L 3G1, Canada}
\affiliation[46]{Instituto de Astrof\'{i}sica de Andaluc\'{i}a (CSIC), Glorieta de la Astronom\'{i}a, s/n, E-18008 Granada, Spain}
\affiliation[47]{Kavli Institute for Particle Astrophysics and Cosmology, 382 Via Pueblo Mall Stanford, CA 94305-4060, USA}
\affiliation[48]{Department of Physics and Astronomy, Sejong University, Seoul, 143-747, Korea}
\affiliation[49]{CIEMAT, Avenida Complutense 40, E-28040 Madrid, Spain}
\affiliation[50]{SLAC National Accelerator Laboratory, Menlo Park, CA 94025, USA}
\affiliation[51]{Kavli Institute for Particle Astrophysics and Cosmology and Department of Physics, Stanford University, Stanford, CA 94305, USA}
\affiliation[52]{Space Telescope Science Institute, 3700 San Martin Drive, Baltimore, MD 21218, USA}
\affiliation[53]{Department of Physics, University of Michigan, Ann Arbor, MI 48109, USA}
\affiliation[54]{University of Michigan, Ann Arbor, MI 48109, USA}
\affiliation[55]{Physics and Astronomy Department, Stony Brook University, Stony Brook, NY 11794}
\affiliation[56]{Department of Physics \& Astronomy, Ohio University, Athens, OH 45701, USA}
\affiliation[57]{School of Earth and Space Exploration, Arizona State University, 781 Terrace Mall, Tempe, AZ 85287, U.S.A.}
\affiliation[58]{Instituto de F\'isica, Pontificia Universidad Cat\'olica de Valpara\'iso, Casilla 4059, Valpara\'iso, Chile}
\affiliation[59]{Department of Astronomy, Cornell University, Ithaca, NY, 14853, USA}
\affiliation[60]{NASA Goddard Space Flight Center, 8800 Greenbelt Road, Greenbelt MD 20771, USA}
\affiliation[61]{National Astronomical Observatories, Chinese Academy of Sciences, A20 Datun Rd., Chaoyang District, Beijing, 100012, P.R. China}

%% file: acknowledgements.tex
We would like to thank Bruce Partridge for helpful discussions in preparing this manuscript.

JK acknowledges support from NSF grants AST-2307727 and AST-2153201.
NS is supported by the Office of Science Graduate Student Research (SCGSR) program administered by the Oak Ridge Institute for Science and Education for the DOE under contract number DE‐SC0014664. MM acknowledges support from NSF grants AST-2307727 and  AST-2153201 and NASA grant 21-ATP21-0145. 
SF is supported by Lawrence Berkeley National Laboratory and the Director, Office of Science, Office of High Energy Physics of the U.S. Department of Energy under Contract No.\ DE-AC02-05CH11231.
GSF acknowledges support through the Isaac Newton Studentship and the Helen Stone Scholarship at the University of Cambridge. GSF and BDS acknowledge support from the European Research Council (ERC) under the European Union’s Horizon 2020 research and innovation programme (Grant agreement No. 851274).
EC acknowledges support from the European Research Council (ERC) under the European Union’s Horizon 2020 research and innovation programme (Grant agreement No. 849169).
GAM is part of Fermi Research Alliance, LLC under Contract No. DE-AC02-07CH11359 with the U.S. Department of Energy, Office of Science, Office of High Energy Physics. 
KM acknowledges support from the National Research Foundation of South Africa.
CS acknowledges support from the Agencia Nacional de Investigaci\'on y Desarrollo (ANID) through Basal project FB210003.
OD acknowledges support from a SNSF Eccellenza Professorial Fellowship (No. 186879).
JD acknowledges support from NSF award AST-2108126.
CEV received the support of a fellowship from “la Caixa” Foundation (ID 100010434). The fellowship code is LCF/BQ/EU22/11930099.
IAC acknowledges support from Fundaci\'on Mauricio y Carlota Botton and the Cambridge International Trust.
This research has made use of NASA's Astrophysics Data System and the arXiv preprint server.

Support for ACT was through the U.S.~National Science Foundation through awards AST-0408698, AST-0965625, and AST-1440226 for the ACT project, as well as awards PHY-0355328, PHY-0855887 and PHY-1214379. Funding was also provided by Princeton University, the University of Pennsylvania, and a Canada Foundation for Innovation (CFI) award to UBC. ACT operated in the Parque Astron\'omico Atacama in northern Chile under the auspices of the Agencia Nacional de Investigaci\'on y Desarrollo (ANID). The development of multichroic detectors and lenses was supported by NASA grants NNX13AE56G and NNX14AB58G. Detector research at NIST was supported by the NIST Innovations in Measurement Science program. Computing for ACT was performed using the Princeton Research Computing resources at Princeton University, the National Energy Research Scientific Computing Center (NERSC), and the Niagara supercomputer at the SciNet HPC Consortium. SciNet is funded by the CFI under the auspices of Compute Canada, the Government of Ontario, the Ontario Research Fund–Research Excellence, and the University of Toronto. We thank the Republic of Chile for hosting ACT in the northern Atacama, and the local indigenous Licanantay communities whom we follow in observing and learning from the night sky.

This material is based upon work supported by the U.S. Department of Energy (DOE), Office of Science, Office of High-Energy Physics, under Contract No. DE-AC02-05CH11231, and by the National Energy Research Scientific Computing Center, a DOE Office of Science User Facility under the same contract. Additional support for DESI was provided by the U.S. National Science Foundation (NSF), Division of Astronomical Sciences under Contract No. AST-0950945 to the NSF's National Optical-Infrared Astronomy Research Laboratory; the Science and Technologies Facilities Council of the United Kingdom; the Gordon and Betty Moore Foundation; the Heising-Simons Foundation; the French Alternative Energies and Atomic Energy Commission (CEA); the National Council of Science and Technology of Mexico (CONACYT); the Ministry of Science and Innovation of Spain (MICINN), and by the DESI Member Institutions: \url{https://www.desi.lbl.gov/collaborating-institutions}.

The DESI Legacy Imaging Surveys consist of three individual and complementary projects: the Dark Energy Camera Legacy Survey (DECaLS), the Beijing-Arizona Sky Survey (BASS), and the Mayall $z$-band Legacy Survey (MzLS). DECaLS, BASS and MzLS together include data obtained, respectively, at the Blanco telescope, Cerro Tololo Inter-American Observatory, NSF's NOIRLab; the Bok telescope, Steward Observatory, University of Arizona; and the Mayall telescope, Kitt Peak National Observatory, NOIRLab. NOIRLab is operated by the Association of Universities for Research in Astronomy (AURA) under a cooperative agreement with the National Science Foundation. Pipeline processing and analyses of the data were supported by NOIRLab and the Lawrence Berkeley National Laboratory. Legacy Surveys also uses data products from the Near-Earth Object Wide-field Infrared Survey Explorer (NEOWISE), a project of the Jet Propulsion Laboratory/California Institute of Technology, funded by the National Aeronautics and Space Administration. Legacy Surveys was supported by: the Director, Office of Science, Office of High Energy Physics of the U.S. Department of Energy; the National Energy Research Scientific Computing Center, a DOE Office of Science User Facility; the U.S. National Science Foundation, Division of Astronomical Sciences; the National Astronomical Observatories of China, the Chinese Academy of Sciences and the Chinese National Natural Science Foundation. LBNL is managed by the Regents of the University of California under contract to the U.S. Department of Energy. The complete acknowledgments can be found at \url{https://www.legacysurvey.org/}.

Any opinions, findings, and conclusions or recommendations expressed in this material are those of the author(s) and do not necessarily reflect the views of the U. S. National Science Foundation, the U. S. Department of Energy, or any of the listed funding agencies.

The authors are honored to be permitted to conduct scientific research on Iolkam Du'ag (Kitt Peak), a mountain with particular significance to the Tohono O'odham Nation.